\documentclass[aps,prd,preprint,superscriptaddress,nofootinbib,showpacs,longbibliography]{revtex4-1}

\usepackage{graphicx}
\usepackage{dcolumn}
\usepackage{bm}
\usepackage{color}%
\usepackage[english]{babel}
\usepackage{amsmath,amssymb}
\usepackage{amsfonts}
\usepackage{booktabs}
\usepackage{multirow}
\usepackage{mathrsfs}
\usepackage{epsfig}
\usepackage{tabularx}
\usepackage{array}
\usepackage{mathtools}
\usepackage{enumerate}
\usepackage{booktabs}
\usepackage{siunitx}
\usepackage{colortbl}
\usepackage{xcolor}
\usepackage[%
  colorlinks=true,
  urlcolor=blue,
  linkcolor=blue,
  citecolor=blue
]{hyperref}
\allowdisplaybreaks[4]  

\newcommand{\pppp}{\rm 4\gamma} 
 
\newcommand{\ppll}{\rm 2\gamma~2 \ell}
\newcommand{\ppgg}{\rm 2\gamma~2 g}
\newcommand{\lambdau}{\Lambda_\mathcal{U}}
\newcommand{\up}{\mathcal{U}}
\newcommand{\ra}{\rightarrow}
\newcommand{\oo}{\mathcal{O}}
\newcommand{\uu}{\mathcal{U}}
\newcommand{\ad}{A_{d_\mathcal{U}}}
\newcommand{\du}{d_\mathcal{U}}
\newcommand{\eq}{Eq.}

\AtBeginDocument{
\heavyrulewidth=.08em
\lightrulewidth=.05em
\cmidrulewidth=.03em
\belowrulesep=.65ex
\belowbottomsep=0pt
\aboverulesep=.4ex
\abovetopsep=0pt
\cmidrulesep=\doublerulesep
\cmidrulekern=.5em
\defaultaddspace=.5em
}
\begin{document}	
	
	\title{Scalar Unparticle Signals at the LHC}

	\newcommand{\metu}{Department of Physics, Middle East Technical University, Ankara 06800, Turkey}
    \newcommand{\usa}{Physics Department, University of California, Santa Barbara, CA 93106-9530, USA}
	\author{T.M.~Aliev}\email[]{taliev@metu.edu.tr}  \affiliation{ \metu }
	\author{ S.~Bilmi\c{s}} \email[]{sbilmis@metu.edu.tr} \affiliation{ \metu }
	\author{ M.~Solmaz}\email[]{msolmaz@hep.ucsb.edu} \affiliation{ \usa }
	\author{I.~Turan}\email[]{ituran@metu.edu.tr}  \affiliation{ \metu }
	
	
	\date{\today}
	
	\begin{abstract}
If the scale invariance exists in nature, the so-called unparticle physics may become part of reality. The only way to refute or confirm this idea is through the experiments one of which is the Large Hadron Collider (LHC). One of the peculiar properties of the unparticle stuff is that it gives striking multi-photon signals which has been studied through only the unparticle self-interactions.  By considering not only the self-interactions of unparticles but also all the other possible contributions, which are dominant,  a detailed study of the processes, within a scalar unparticle scenario,  $pp\to 4\gamma$, $pp\to 2\gamma 2g$, $pp\to 2\gamma 2\ell$, $pp\to 4e$, $pp\to 4\mu$ and $pp\to 2e2\mu$ at $\sqrt{s}=14$ TeV at the LHC is carried out. We use basic selection cuts and analyze various distributions to discriminate the signals over the Standard Model backgrounds and discuss what seems to be the most likely channel among the above ones for an indirect manifestation of unparticle effects. We follow a new approach to tackle the issue with the three-point correlation function for the scalar unparticle self-interactions. We also obtain the exclusion region in the unparticle parameter space from the available two-photon data of the LHC.
	\end{abstract}
	
	
	\maketitle
	
	
	\tableofcontents
	
	\section{Introduction}
	After the discovery of the Higgs particle at the Large Hadron Collider (LHC) at CERN~\cite{Aad:2012tfa,Chatrchyan:2012xdj}, the particle content of the Standard Model (SM) has finally been completed after so many years of desperate search. Despite the fact that SM is extremely successful in describing all existing experimental data, it has  still been lacking mechanisms to explain some unsolved problems. For example, it could not explain neutrinos to be massive, does not include gravity, has no dark matter candidate, subsumes the so-called hierarchy problem, etc.
	Having the Higgs particle at hand, the following three distinct directions will shape the search programs in the current and upcoming experiments:
\begin{enumerate}[(a)]
	\item Making precise measurements of the Higgs decay channels, the Yukawa couplings, etc.
	\item Improving the precision to measure the properties of the SM particles as well as the electroweak precision parameters such as the electroweak mixing angle, $W$ boson mass, asymmetries, etc.
	\item Searching for new physics beyond the SM, among the vast list of which are low-scale supersymmetry, extra dimensions, and the so-called unparticle physics which originates from entirely different standpoint. 
\end{enumerate}
	
	About a decade ago, unparticle physics as a beyond scenario has been introduced in~\cite{Georgi:2007ek,Georgi:2007si} based on the low energy manifestation of a non-trivial scale invariant effective field theory. In this content, as the simplest choice, a new scalar field (called scalar unparticle $\mathcal{U}$) which is a singlet under $SU(2)_L$ group can couple to photons and gluons directly through higher dimensional operators with a cut-off scale $\lambdau$ below which interpolating fields emerge with some non-integral scaling dimension $\du$. The scenario involves rich phenomenology and predicts the existence of scalar unparticle self-interactions \cite{Feng:2008ae,Georgi:2009xq,Bergstrom:2009iz,Cheung:2007zza,Cheung:2007ap,Cheung:2008xu}, which could give unusually large effects in gluon fusion processes. For example, the $g g \ra \up \ra \gamma \gamma$ process  leads to enhancement of signals in the Higgs decay channels and the self-interactions of unparticles give rise to signals with different four particle states such as four photons, two photons + two gluons, two photons + two leptons, and four charged leptons. It is interesting that the four photon signal is practically background free and therefore can play a critical role in the discovery of unparticles (for more details see~\cite{Bergstrom:2009iz,Aliev:2009sd}). It has also been shown that in addition to the contribution to some of these processes through the scalar unparticle self-interactions, there are other single and double unparticle exchange diagrams, making significant contributions (even dominating) to these signals \cite{Aliev:2009sd}.  Hence it is essential to do a complete study of such signals including all contributions.

	In the present work, we extend the calculations presented in~\cite{Aliev:2009sd} for the processes $p p \ra \pppp$, $p p \ra \ppgg$  to the LHC energy $\sqrt{s} = 14$ TeV by making a simulation including basic detector effects, as well as analyzing the other processes with the final states $\ppll$, $e^+e^- e^+e^-$,  $\mu^+\mu^- \mu^+ \mu^-$, and $e^+e^- \mu^+ \mu^-$ at the LHC.
	
	The work is organized as follows. In section~\ref{Sec:model} we briefly describe the elements of the unparticle theory, present the specific couplings necessary for our calculations. Unparticle self-interactions and how we treat the vertex function are given in section~\ref{Sec:self}. Section~\ref{Sec:madgraph} covers some details of the scalar unparticle model implementation to {\tt MadGraph5}.   Section~\ref{Sec:num} is devoted to the numerical analysis of the processes with different four particles in the final states. In Section~\ref{Sec:conc}, we give summary of our work.

	\section{Theoretical Framework}\label{Sec:model}
	The basic idea of the unparticle theory is the existence of scale invariant hidden sector at high energy $\lambdau$. Below the $\lambdau$ scale,  unparticle physics manifest as interpolating field $\oo$ having various scaling dimensions and Lorentz structure. One of characteristic property of unparticle operator is that it has a continuous spectral density
	\begin{equation}
	\rho(p^2) = \ad\oo(p^0)\oo(p^2)(p^2)^{\du-2}~,
	\label{density}
	\end{equation}
	where $\du$ is the scale dimension parameter and the factor $\ad$ is determined as;
	
	\begin{equation}
	\ad = \frac{16 \pi^{5/2}}{(2 \pi)^{3/2}} \frac{\Gamma(\du+1/2)}{\Gamma(\du-1)\Gamma(2 \du)}~.
	\end{equation}
	
	From this expression, it follows that when $\du \ra 1$, \eq ~\ref{density} reduces to the massless particle phase space. For this reason, one can suggest that unparticle behaves like a collection of $\du$ massless fields. 
	In the rest of the paper, we restrict ourselves by considering only scalar unparticle.	The form of propagator for scalar particle is obtained in~\cite{Cheung:2007zza,*Cheung:2007ap,*Cheung:2008xu}
	
	\begin{equation}
	\Delta_f = \frac{\ad}{2 \sin(\pi\du)} \frac{i e^{i \phi}}{(|p|^2+i\epsilon)^{2-\du}}
	\label{delta}
	\end{equation}
	
	The phase $\phi$ is defined as $\phi = Arg(-p^2)^{\du}$. It should be noted that the phase is nonzero in $s$-channel, while in $t$ and $u$ channels it is equal to zero. For the  scalar operator, the unitarity condition leads to $\du \ge 1$ \cite{Grinstein:2008qk}.
	Unparticle operators can interact with the SM particles via exchange of heavy particle with mass $M$. After integrating out the heavy degrees of freedom, a series of effective operators describing the interaction of the SM particles with unparticles at low energy are obtained. The operators describing the interactions for scalar unparticle with the SM particles are; 
	 \begin{eqnarray}
	 \lambda_0^\prime \frac{1}{\lambdau^{\du-1}} \bar{f}f\oo, \\ \nonumber
	 \lambda_0^{\prime\prime} \frac{1}{\lambdau^{\du-1}} \bar{f}i\gamma_5f\oo, \\ \nonumber
	 \lambda_0 \frac{1}{\lambdau^{\du}} G_{\alpha \beta} G^{\alpha \beta} \oo~.
	 \end{eqnarray}
	 
	 The Feynman rules for the scalar unparticle operators with the $g g$ and $\gamma \gamma$ are
	 \begin{equation}
	 4i \lambda_{g,\gamma}^0 \frac1{\lambdau}(-p_1p_2g_{\mu \nu}+p_{1\nu}+p_{2\mu})~.
	 \end{equation}
     For the calculation of the signals at hand the following two- and three-point correlation functions need to be evaluated \cite{Feng:2008ae,Bergstrom:2009iz}
	 
	 \begin{equation}
	 < 0 | \oo_\uu(x) \oo_\uu^\dagger (0)|0 > = \int \frac{d^4 p}{(2 \pi^4)} e^{-ipx}\rho_\uu(p^2)~,
	 \end{equation}

	 \begin{eqnarray}
	 	 <0|\oo_\uu(p_1) \oo_\uu(p_2) \oo_\uu^\dagger(p_1+p_2)|0> && = C_d \int \frac{d^4 q}{(2\pi)^4} \{ [-q^2-i\epsilon][-(p_1-q)^2 -i\epsilon] \nonumber \\ 
	 	 && \times[-(p_2+q)^2 -i\epsilon] \}^{\du/2-2} \nonumber \\
	 	 && = -i (-1)^n C_d(\frac{1}{s})^{n-2}F_y (\frac{p_1^2}{s},\frac{p_2^2}{s})
	 	 \label{threepoint}
	 \end{eqnarray}
	 where $n = 6(1-\du/4)$ and $s=(p_1+p_2)^2$. The three-point correlation function is	
	 \begin{eqnarray}
	 F_y (\frac{p_1^2}{s},\frac{p_2^2}{s}) = \frac{\Gamma(n-2)}{16 \pi^2 [\Gamma(n/2)]^3} \int_{0}^{1} dx_1dx_2dx_3\left(x_1 x_2 x_3\right)^{\frac{n}{3}-1} \delta(x_1+x_2+x_3-1) \left(\frac{1}{\Delta}\right)^{n-2}
	 \end{eqnarray}
	 with $\Delta = x_1 x_2\, p_1^2/s + x_1 x_3\, p_2^2/s + x_2x_3$.
	We take $\lambda_{g,\gamma}^0 = 1$ and $\lambda_0^\prime = \sqrt{2\pi}/e$ which follows from the naturalness requirement. The relevant part of the unparticle model has been implemented in $\tt MadGraph\, 5$ \cite{Alwall:2011uj} package program  in the UFO format . Some of the details of the implementation are summarized below.

	\subsection{Unparticle Self-Interaction\label{Sec:self}}
	
	The most striking feature of the unparticle scenario is that it enables three-point vertices where a scalar unparticle couples to two other unparticles of the same type and the vertex factor is not of a typical tree level form. Hence it requires special attention.   Some promising processes such as $pp \to \gamma \gamma \gamma \gamma$, $pp \to \gamma \gamma gg$, $pp \to \gamma \gamma \ell\ell$, etc. could originate from scalar unparticle self-interactions where the factor $C_d$ in \eq ~\ref{threepoint} is indeed free at first. However, see the discussion in \cite{Aliev:2009sd} about various phenomenological bounds on $C_d$ as well as in \cite{Delgado:2009vb,*Caracciolo:2009bx} for theoretical considerations.
	
	Especially, the difficulty behind the computation of the complicated function of $F_{y}$ in the simulations is related to long hours of CPU time due the integrals involved in $F_y$. has led us to pursue a relatively simpler workaround. In this approach, one would not only avoid the time-consuming computation but also make the complete model feasible to  implement on event generator programs. For that reason, we decided to make two-dimensional fitting for $F_{y}$ function. 
	
	For each value of the scaling dimension parameter, $d_{\cal U}= \{1.1, 1.2,...., 1.9\}$, the function $F_y$ has been evaluated via {\tt Mathematica} with a statistically high number of two-dimensional data grids. Afterward, the tabulated dataset for each $d_{\cal U}$ value has been inputted to fitting package of {\tt Matlab} to get polynomial functional forms. Several plots are obtained to check if the fitting results are fairly convincing and some are shown in Fig.~\ref{Fig:fit}. The explicit forms of the fitted function $F_y$ are given for various $\du$ values in Appendix \ref{func}.
	
	\begin{figure}[h]
\hspace{-0.5cm}
		\begin{tabular}{@{}c@{\hskip 0.2in}c}    
		\hspace*{-0.9cm}	\includegraphics[width=3.4in,height=3.4in]{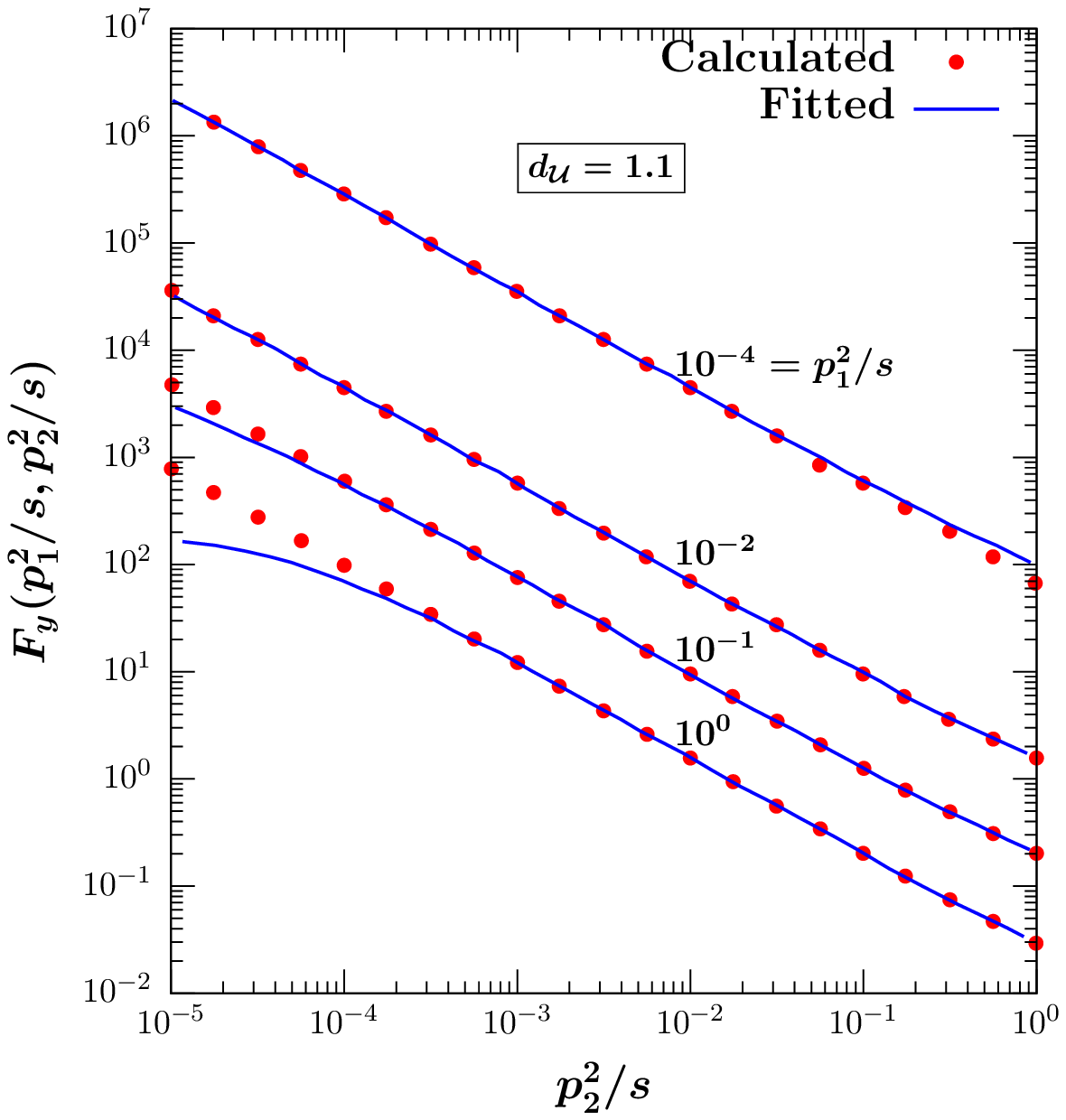} &
		\hspace*{-0.9cm}	\includegraphics[width=3.4in,height=3.4in]{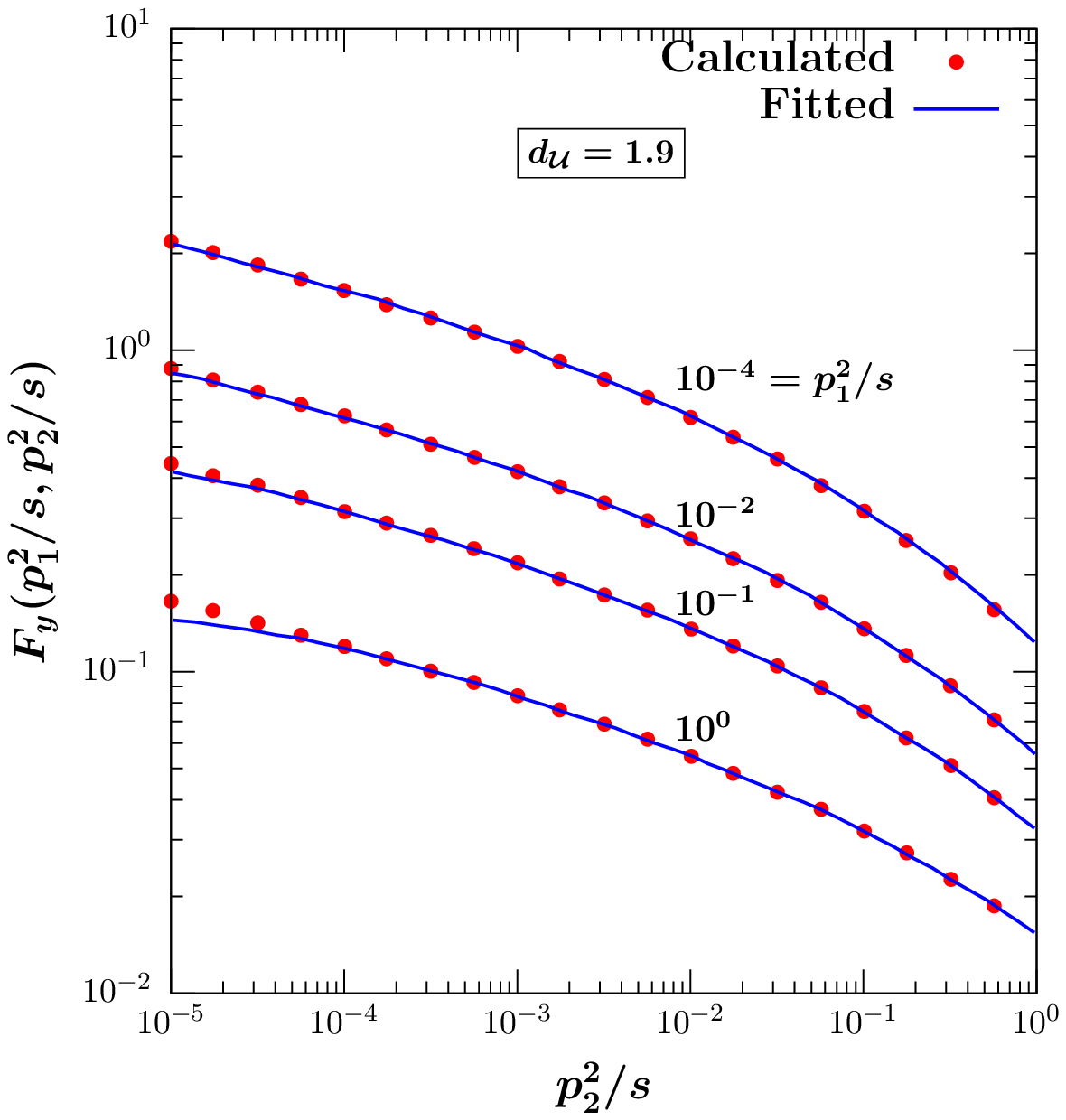} \\
			\includegraphics[width=3.15in,height=3.4in]{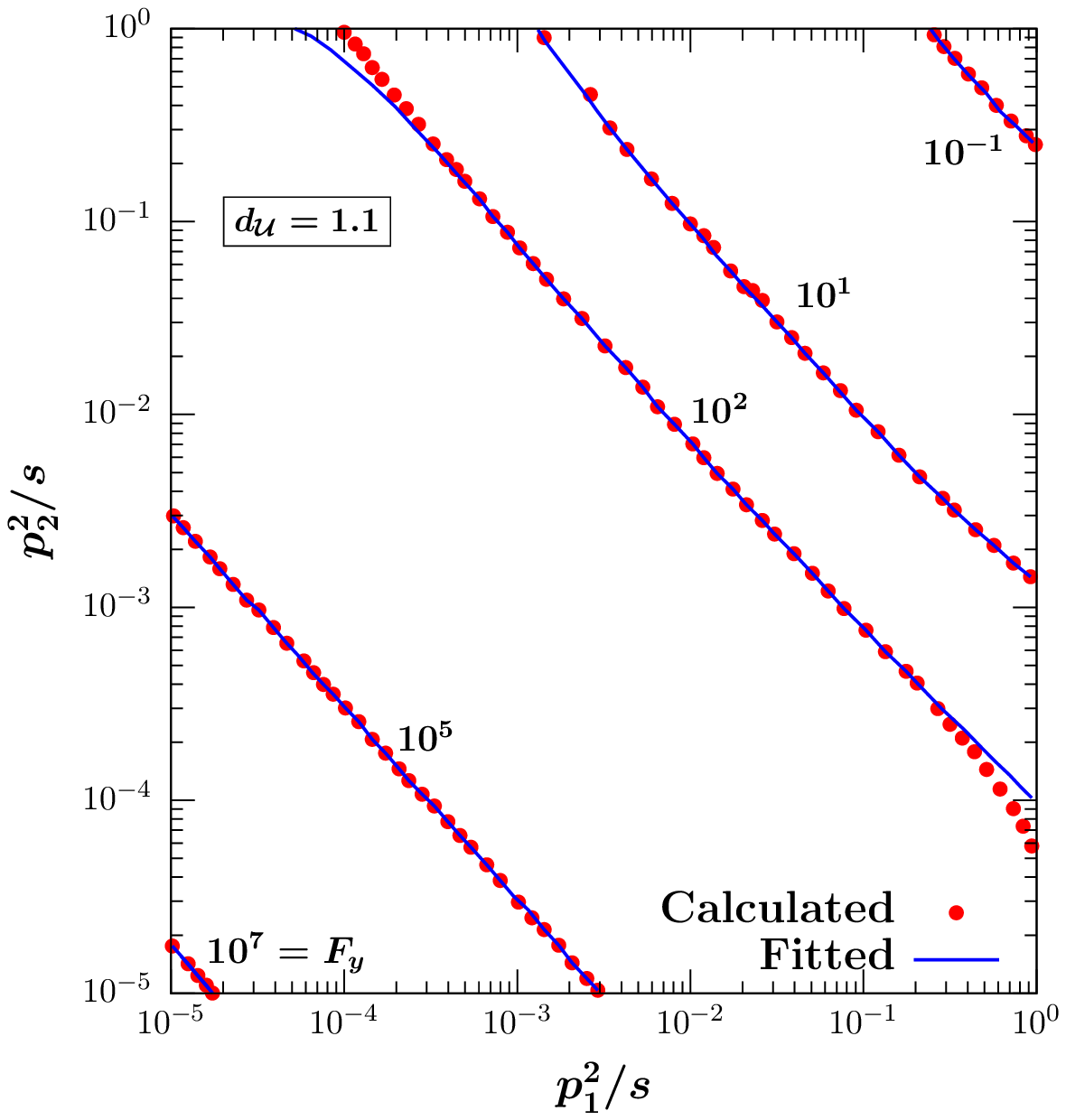} &
			\includegraphics[width=3.15in,height=3.4in]{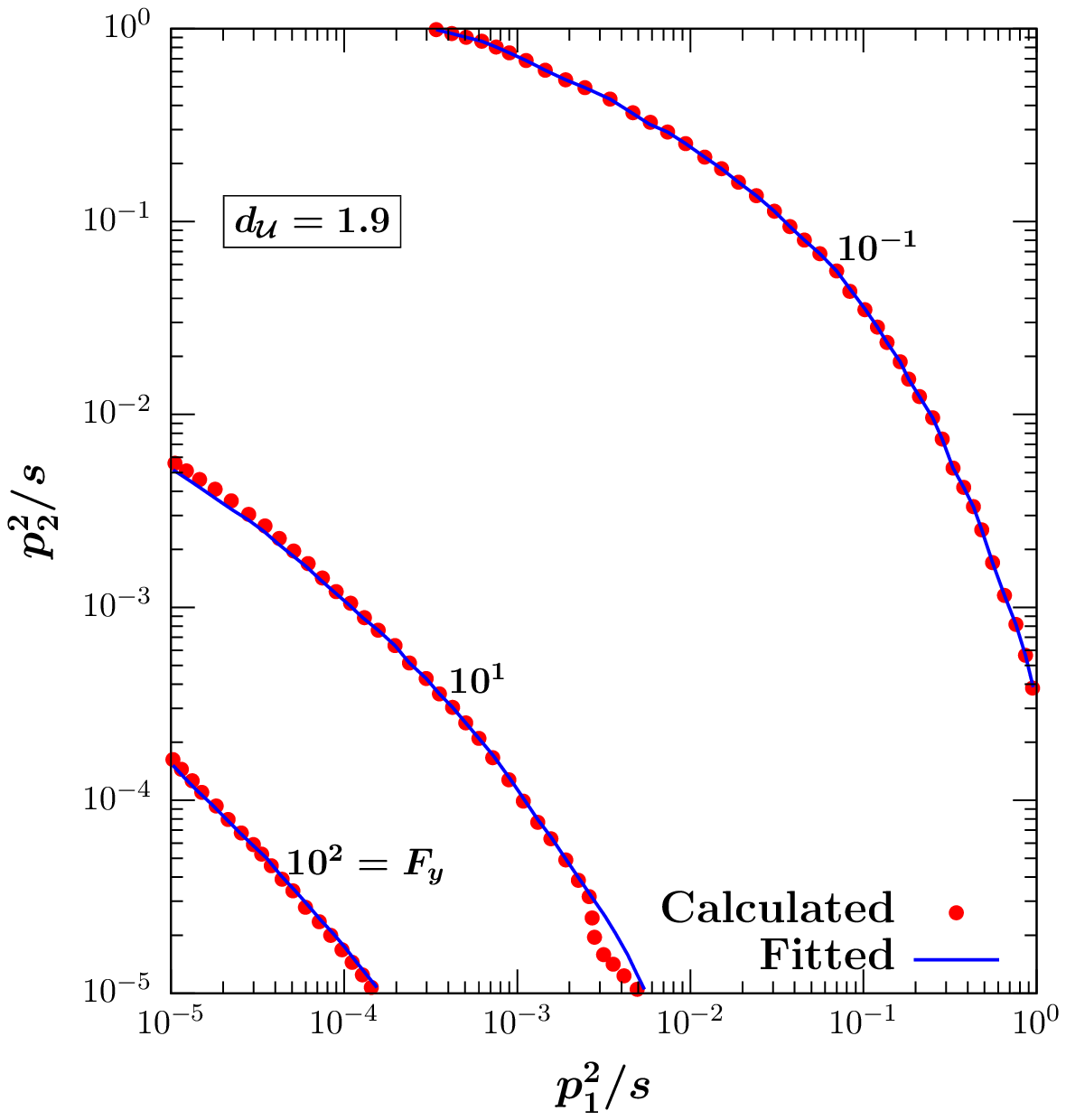} \\
		\end{tabular}
		\caption{\label{Fig:fit} In the upper row, the three-point correlation function $F_y(p_1^2/s,p_2^2/s)$ is plotted as a function of $p_2^2/s$ for various $p_1^2/s$ values by using both the exact integral form and the polynomial fit functions. In the second row, various values of $F_y$ is depicted in the $(p_1^2/s,p_2^2/s)$.}
	\end{figure}

	\subsection{Implementing the Unparticle Model in {\tt MadGraph\,5}\label{Sec:madgraph}}
	
	Implementation of the unparticle model in the event generator programs intended for three-level calculations has not been an easy task to achieve due to the structure of the model, such as non-trivial scalar propagator expression and three-point correlation function.
	
	An event generator program that would offer a vast flexibility of applying non-straightforward principles would be the best choice and {\tt MadGraph\, 5}  fits the purpose. 	Another reason why {\tt MadGraph\, 5} has been chosen is that new physics models can be defined as UFO format \cite{Degrande:2011ua}. There are a couple of advantages in introducing unparticle model as UFO file, one of which is that one may freely write down any Lorentz expression for an arbitrary vertex. In addition, UFO also allows users to define effective vertices with no constraint on the number of particles in each vertex. These features have been employed to define the vertices in the model for  further analysis of the signals of the unparticle model.
	
	The unparticles are defined as massless scalars at Lagrangian level with the {\tt FeynRules} interface \cite{Alloul:2013bka}. The additional parameters were also attributed to the model, such as $d_{\cal U}$ and $\Lambda_{\cal U}$ and some other coupling constants. In the end, {\tt FeynRules} package produces the UFO file of the model containing all the information regarding parameters, couplings, vertices and Lorentz expressions of each vertex.
	
	Further modifications in the  UFO model file are needed to define the unparticle model properly. Then,  processes occurring within the scalar unparticle self-interactions have been introduced to the unparticle model file by setting new effective vertices with two incoming and four outgoing particles, namely $gg/q\bar{q} \to 4\gamma, 2\gamma 2g, 2\gamma 2\ell, 2e 2\mu$. Couplings for these vertices were also added respectively. Moreover, the standalone use of ALOHA \cite{deAquino:2011ub} package led us to scrutinize the Fortran subroutines that belong to the unparticle model, evaluating the amplitude of each Feynman diagram. In this way, we could embed the unparticle scalar propagator and the three-point correlation function into the corresponding subroutines to get the final model file.

	
	\subsection{Bounds in the $(\du,\lambdau)$ Plane\label{Sec:bound}}
	
	Before concentrating on various signals within the unparticle scenario at a center of mass energy of 14 TeV at LHC, let us check the status of the model in the light of the available data. One of the relevant constraints could come from the measurement of an isolated photon pair by the CMS collaboration at 7 TeV \cite{Chatrchyan:2011qt} with a data sample corresponding to an integrated luminosity of 36 pb$^{-1}$ and isolated photons are required to have transverse energies $E_T>23$ GeV and $E_T>20$ GeV, respectively. 
	
	The experimental analysis is performed in two different pseudorapidity regions; one with $|\eta|<1.44$ and the other one $|\eta|<2.5$ but excluding the region $1.44<|\eta|<1.57$. The cone size between the photons is assumed $\Delta R >0.45$. The background events could be like Drell-Yan events with two misidentified electrons as photons, or photon+jet, or multijet events where photons come off hadronic decays. The leading contributions are $q\bar{q}$ annihilation to a diphoton pair, diphoton pair through a gluon fusion, quark-gluon scattering into a diphoton and jet. The results for the integrated diphoton cross sections are \cite{Chatrchyan:2011qt}
\begin{equation*}
\label{equ:totxxsec}
\begin{array}{rcl}
\sigma^{\text{exp}}(pp\rightarrow\gamma\gamma)|_{|\eta|< 1.44} &=& 31.0 \ \ \pm 1.8
\   (\text{stat.}) \ \ \mbox{ }^{ +2.0}_{ -2.1} \   (\text{syst.}) \ \ \pm 1.2 \  (\text{lumi.}) \ \ \ \text{pb}~,\\
~\\
\sigma^{\text{exp}}(pp\rightarrow\gamma\gamma)|_{|\eta|< 2.50} &=& 62.4 \ \ \pm 3.6
\   (\text{stat.}) \ \ \mbox{ }^{ +5.3}_{ -5.8} \   (\text{syst.}) \ \ \pm 2.5 \  (\text{lumi.}) \ \ \ \text{pb}.\\
\end{array}
\end{equation*}
While the theoretical calculation within the Standard Model are computed as \cite{Chatrchyan:2011qt}
\begin{equation*}
\label{equ:tottxsec}
\begin{array}{rcl}
\sigma^{\text{SM}}(pp\rightarrow\gamma\gamma)|_{|\eta| < 1.44} & = & 27.3 \ \ \mbox{ }^{ +3.0}_{ -2.2} \ \ \text{(scales)} \ \ \pm 1.1 \ \ \text{(PDF)} \ \ \text{pb}~,\\
~\\
\sigma^{\text{SM}}(pp\rightarrow\gamma\gamma)|_{|\eta| < 2.50} & = & 52.7 \ \ \mbox{ }^{ +5.8}_{ -4.2} \ \ \text{(scales)} \ \ \pm 2.0 \ \ \text{(PDF)} \ \ \text{pb}.\\
\end{array}
\end{equation*}
Here $|\eta|<1.44$ and $|\eta|<2.5$ are the pseudorapidity regions as described above. Once can see from these numbers that the measurements are consistent with the SM predictions by taking the experimental and theoretical uncertainties into account. 

\begin{figure}[h]
			\begin{center}
				\hspace*{-1.8cm}
				$\begin{array}{c}
				\includegraphics[width=4.5in,height=4.1in]{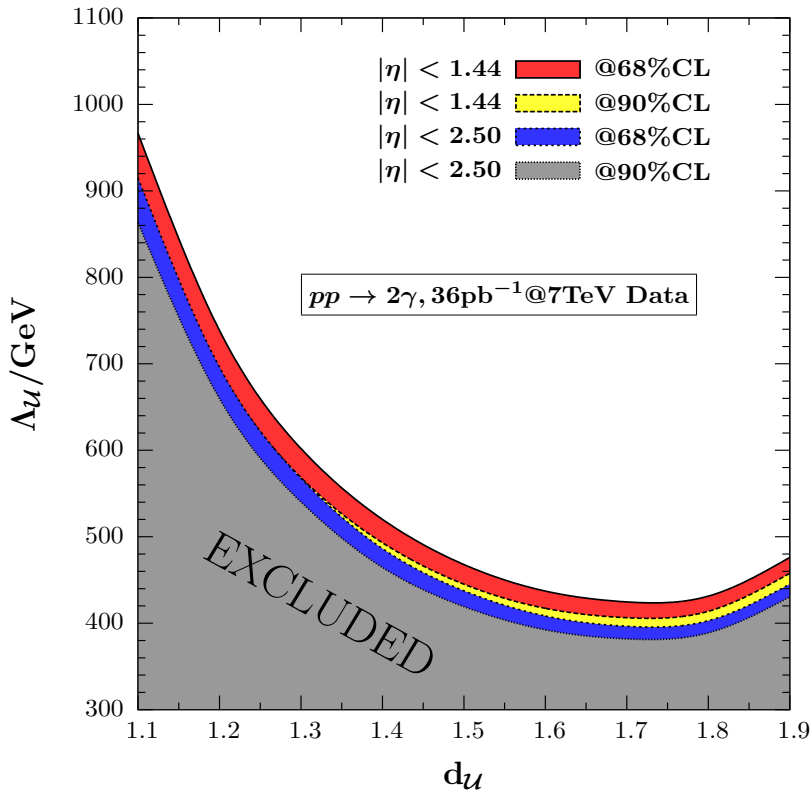} 
				\end{array}$
			\end{center}
			\vskip -0.2in
			\caption{Using the $pp\to 2\gamma$ data with $36$ pb$^{-1}$ at 7 TeV \cite{Chatrchyan:2011qt}, the exclusion plot in the $(\du,\lambdau)$ plane is shown at both $68\%$ and $90\%$ CLs. Two different pseudorapidity cuts are shown. Both the experimental and the SM errors are included.}\label{fig:bound}
		\end{figure}

	If one extends the theoretical framework into unparticle scenario, the theory predictions for the above cross sections would get new indirect contributions as the pure unparticle part, $\sigma^{\text{U}}(pp\rightarrow\gamma\gamma)$ and the interference, $\sigma^{\text{int}}(pp\rightarrow\gamma\gamma)$. Using the available room between the experimental and SM values including both the experimental and theoretical errors, one can set limits on the parameters of unparticle model, namely $d_{\cal U}$ and $\Lambda_{\cal U}$. The exclusion limits in the $(d_{\cal U}, \Lambda_{\cal U})$ plane are shown in Fig.~\ref{fig:bound}. We present our results for both pseudorapidity regions, $|\eta|<1.44$ and $|\eta|<2.5$, at 68$\%$ C.L. and 90$\%$ C.L. in each case. The bound on $\Lambda_{\cal U}$ can get as large as 1 TeV for small $d_{\cal U}$ values, but it is smaller for larger $d_{\cal U}$ values. Note that the analysis of the SM part in \cite{Chatrchyan:2011qt} has already been calculated in the next to leading order, but we kept the unparticle contribution as well as the interference in the leading order.

	\section{Numerical Analysis}\label{Sec:num}
	
	A numerical analysis will be done by using various kinematical quantities. Let us briefly explain them. If $\theta$ and $\varphi$ represent the polar and azimuthal angles in the barrel, respectively, the distance between the particle $i$ and particle $j$ of an event can be defined as $\Delta R_{ij} =\sqrt{(\Delta \eta_{ij})^2+(\Delta \varphi_{ij})^2}$ where $\eta_i$ is the pseudorapidity of the particle $i$, defined as $\eta_i=-\ln\left(\tan\frac{\theta_i}{2}\right)$. Here $i$ and $j$ represent any particle in our signals.

	Another kinematical quantity is the invariant mass of the $ij$- particle system and is defined as $m_{ij}=\left( p_i+p_j\right)^2$ where $p_i(p_j)$ is the four-momentum of the particle $i(j)$. This definition can be extended to more than two particles as well. Note also that the broad peaks in the invariant mass distributions do not always correspond to the existence of a new particle and care should be given. 

	        \begin{table}[htb]
	        	\caption{The selection cuts imposed for each channel.}
	        	\label{tab:cut_all}	
	        	\begin{tabular}{c@{\hskip 0.7in}c@{\hskip 0.7in}c@{\hskip 0.7in}c}
	        		\hline
	        		\hline
	        		$pp \rightarrow 4 \gamma$ & $pp \rightarrow 2g 2\gamma$ & $pp \rightarrow 2 \gamma 2\ell$ & $pp \rightarrow 4\ell$ \\ \midrule
	        		
	        		\multirow{2}{*}{$p_T(\gamma)> 30~\rm{GeV}$} & $p_T(\gamma)> 30~\rm{GeV}$ & $p_T(\gamma)> 30~\rm{GeV}$ & \multirow{2}{*}{$p_T(\ell)> 15~\rm{GeV}$} \\  
	        		& $p_T(j) > 30~\rm{GeV}$ &  $p_T(\ell)> 15~\rm{GeV}$ &  \\ \midrule
	        		
	        		\multirow{2}{*}{$ |\rm{\eta}(\gamma)| < 2.44$} & $ |\rm{\eta}(\gamma)| < 2.44$ &$ |\rm{\eta}(\gamma)| < 2.44$  & \multirow{2}{*}{$ |\rm{\eta}(\ell)| < 2.0$} \\
	        		& $ |\rm{\eta}(j)| < 2.44$ & $ |\rm{\eta}(\ell)| < 2.44$ &  \\ \midrule
	        		\multirow{3}{*}{$\rm{\Delta R}(\gamma, \gamma) > 0.4$} & $\rm{\Delta R}(j, j) > 0.4$ & $\rm{\Delta R}(\ell,\ell) > 0.4$ & \multirow{3}{*}{$\rm{\Delta R}(\ell, \ell) > 0.4$} \\
	        		
	        		& $\rm{\Delta R}(\gamma, \gamma) > 0.4$ & $\rm{\Delta R}(\gamma, \gamma) > 0.4$ & \\ 
	        		&$\rm{\Delta R}(j, \gamma) > 0.4$ & $\rm{\Delta R}(\ell,\gamma) > 0.4$ & \\
	        		\hline
	        		\hline
	        	\end{tabular} 
	        \end{table}

	There are two more transverse variables to define. One is the usual transverse momentum of, say, particle $i$, $p_T^i=\sqrt{(p_x^i)^2+(p_y^i)^2}$ if the beam direction is taken along the $z$-axis. For each event, the objects are listed in the  order of decreasing transverse momenta. The other one is the so-called $H_T$ variable, related to $p_T^i$. $H_T$ is defined as the scalar sum of the $p_T^i$ where $i$ could be jet, lepton, or photon as well as missing transverse energy, $/\!\!\!\!E_T$. That is, $H_T=/\!\!\!\!E_T + \sum_i |p_T^i|$. Thus, $H_T$ can be taken as a measure of the overall energy scale of the process.

	The basic cuts applied for each signal are listed in Table \ref{tab:cut_all}. 	All simulations are done by first using {\tt MadGraph\,5} \cite{Alwall:2014hca} to generate partonic events and then {\tt Pythia} \cite{Sjostrand:2006za, *Sjostrand:2014zea} event generator is used for hadronization with parton distribution functions {\tt CTEQ6L1}. The final results are obtained after passing events to ${\tt PGS}$ to simulate the detector limitations. In Table \ref{tab:cs_all}, we list the total cross sections for the channels $4\gamma, 2\gamma2g, 2\gamma 2\ell,$ and $4\ell$ at $\sqrt{s}=14$ TeV center of mass energy for the following values of $d_{\cal U}$ and $\Lambda_{\cal U}$; $(d_{\cal U},\Lambda)=(1.1,1-3\,{\rm TeV}),\, (1.5,1-3\,{\rm TeV}),\, (1.9,1-3\,{\rm TeV})$. The SM cross sections are also included for background comparison. For almost all signals the unparticle cross sections are around two-three orders of magnitude larger than those of the SM for  $\Lambda_{\cal U}=1$ TeV but they become almost the same when $\Lambda_{\cal U}=3$ TeV. A sizable deviation from the background is possible for $\Lambda_{\cal U}$ around 1TeV.

\begin{table}[h]
\caption{The total cross sections (in $pb$) of the signals considered in the study are listed for two  cut off $\Lambda_{\cal U}$ values, 1 TeV and 3 TeV, and various $d$ values. The cross sections for the  Standard Model background are also included for comparison.}
\label{tab:cs_all}
  \begin{tabular}{l@{\hskip 0.38in}c@{\hskip 0.38in}c@{\hskip 0.38in}c@{\hskip 0.38in}c@{\hskip 0.38in}c}
\hline
\hline
 \multicolumn{6}{c}{Cross-section Values (pb)} \\ \cmidrule{1-6}
    &$\Lambda_{\cal U}$ & $d_{\cal U}=1.1$ & $d_{\cal U}=1.5$ & $d_{\cal U}=1.9$ & SM \\ \midrule
  \multirow{2}{*}{$pp\to 4 \gamma$} & $1~{\rm TeV}$ & $9.792 \times 10^{-3}$ & $1.745 \times 10^{-4}$ & $7.665 \times 10^{-4}$ & \multirow{2}{*}{$8.776 \times 10^{-6}$}\\
    & $3~{\rm TeV}$ & $1.077 \times 10^{-5}$ & $1.018 \times 10^{-5}$ & $1.017 \times 10^{-5}$ \\ \midrule
  \multirow{2}{*}{$pp\to 2\gamma 2g$} & $1~{\rm TeV}$ & $5.520 \times 10^{1}$ & $3.010 \times 10^{0}$ & $3.798 \times 10^{0}$ & \multirow{2}{*}{$1.675 \times 10^{-1}$} \\
    & $3~{\rm TeV}$ & $6.166 \times 10^{-1}$ & $1.826 \times 10^{-1}$ & $1.797 \times 10^{-1}$ \\ \midrule
  \multirow{2}{*}{$pp\to 2\gamma 2\ell$ } & $1~{\rm TeV}$ & $8.117 \times 10^{-3}$ & $7.251 \times 10^{-4}$ & $7.716 \times 10^{-4}$ & \multirow{2}{*}{$4.355 \times 10^{-4}$} \\
    & $3~{\rm TeV}$ & $5.060 \times 10^{-4}$ & $4.716 \times 10^{-4}$ & $4.713 \times 10^{-4}$ \\ \midrule
  \multirow{2}{*}{$pp\to 4\ell$} & $1~{\rm TeV}$ & $6.310 \times 10^{-4}$ & $4.422 \times 10^{-5}$ & $5.903 \times 10^{-5}$ & \multirow{2}{*}{$8.586 \times 10^{-6}$} \\
    & $3~{\rm TeV}$ & $1.304 \times 10^{-5}$ & $1.026 \times 10^{-5}$ & $1.021 \times 10^{-5}$ \\
\hline
\hline
\end{tabular}
\end{table}

\subsection{$pp \to \pppp$ Signal}
		
Detecting energetic photons at colliders serves many purposes like testing perturbative QCD \cite{Chatrchyan:2011qt,Aad:2012ans} as well as various commonly used techniques \cite{Lai:1996mg,*Martin:1999ww,*Werlen1999201}. Their better identification becomes critical since they usually form an important background to various exotic signals of the beyond SM scenarios \cite{Orimoto:2015ueg,*Aad:2015bua}. Even though measuring photon pair production signal can be done with some precision, it gets harder as the number of photons increases and the SM prediction gets suppressed. Therefore, multi-photon signals are testing grounds for different scenarios and unparticle would be one of them. 
	
	In this subsection, we will discuss four photon signal within the unparticle framework and then compare with what the SM background looks like. Four-photon event selection requires each photon to have at least $30$ GeV transverse momentum ($p_T$) with a cone separation $\Delta R_{ij}=0.4$ between any two photons. Pseudorapidity $|\eta_i|\le 2.44$ is also required for each photon. They are listed in Table \ref{tab:cs_all}.

		\begin{figure}[h]
			\begin{center}
				\hspace*{-1.8cm}
				$\begin{array}{cc}
				\includegraphics[width=3.5in,height=3.1in]{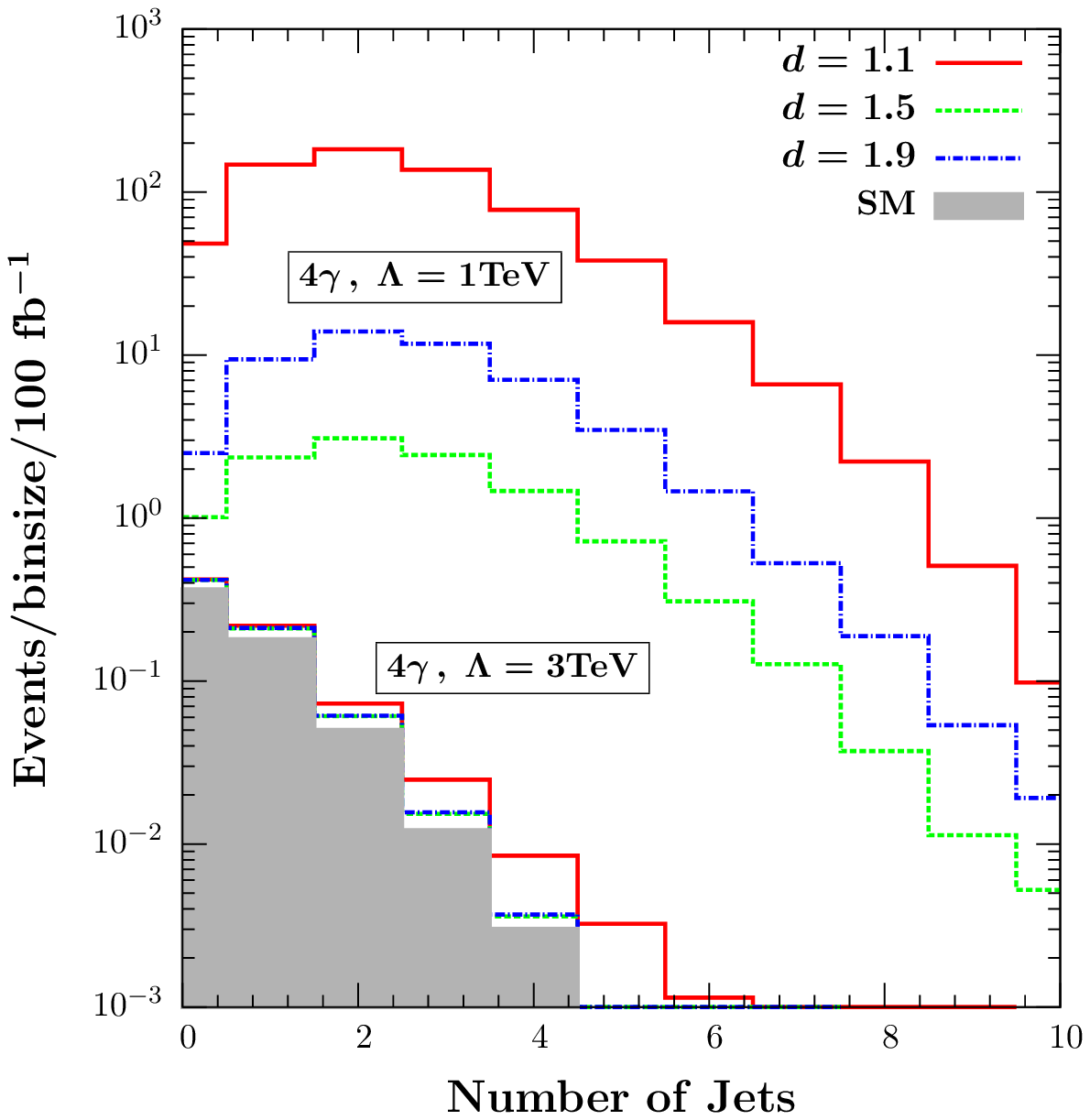} &\hspace*{-1cm}
				\includegraphics[width=3.5in,height=3.1in]{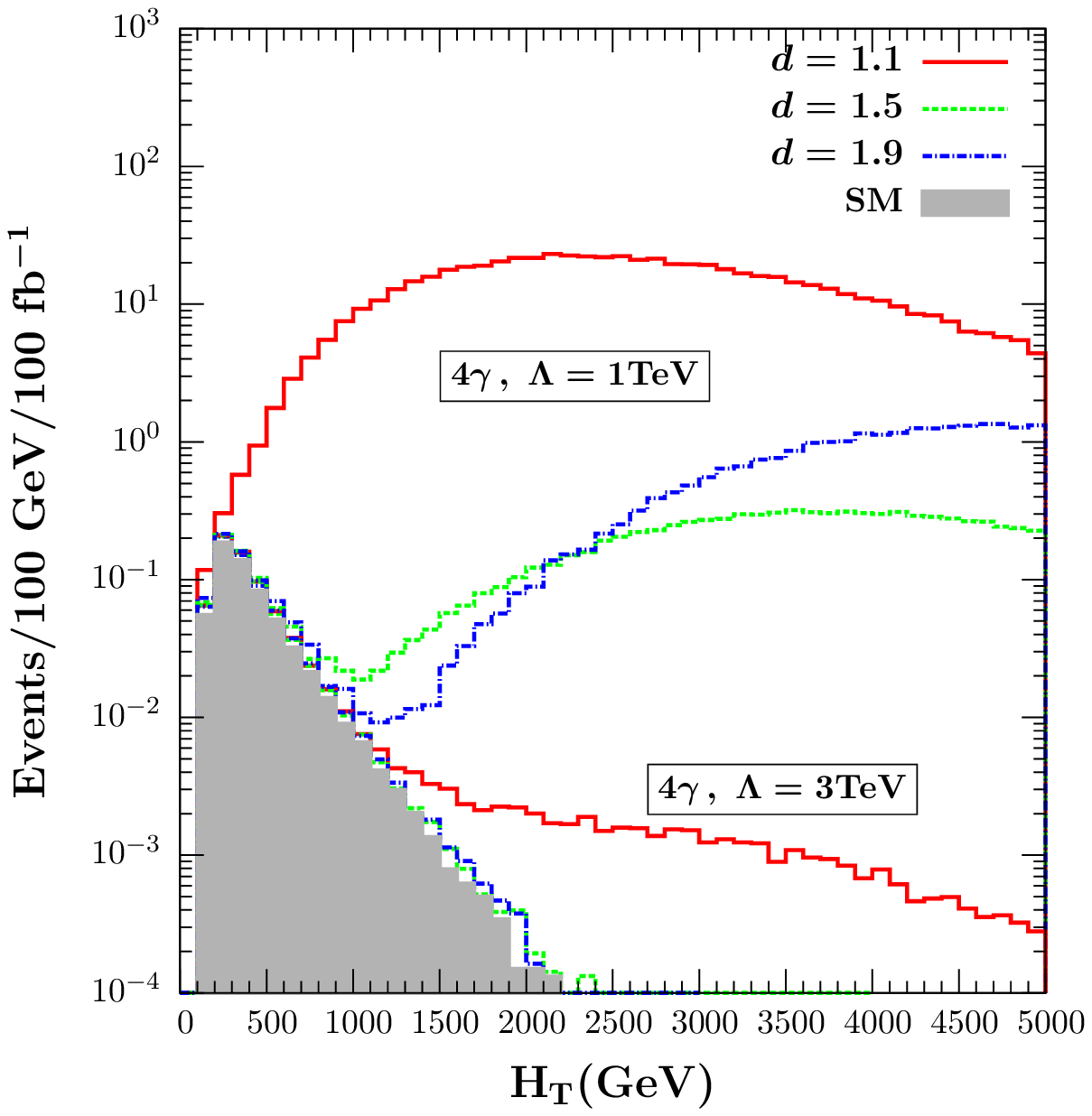} \\ \hspace{0.2cm}
				\includegraphics[width=3.4in,height=3.1in]{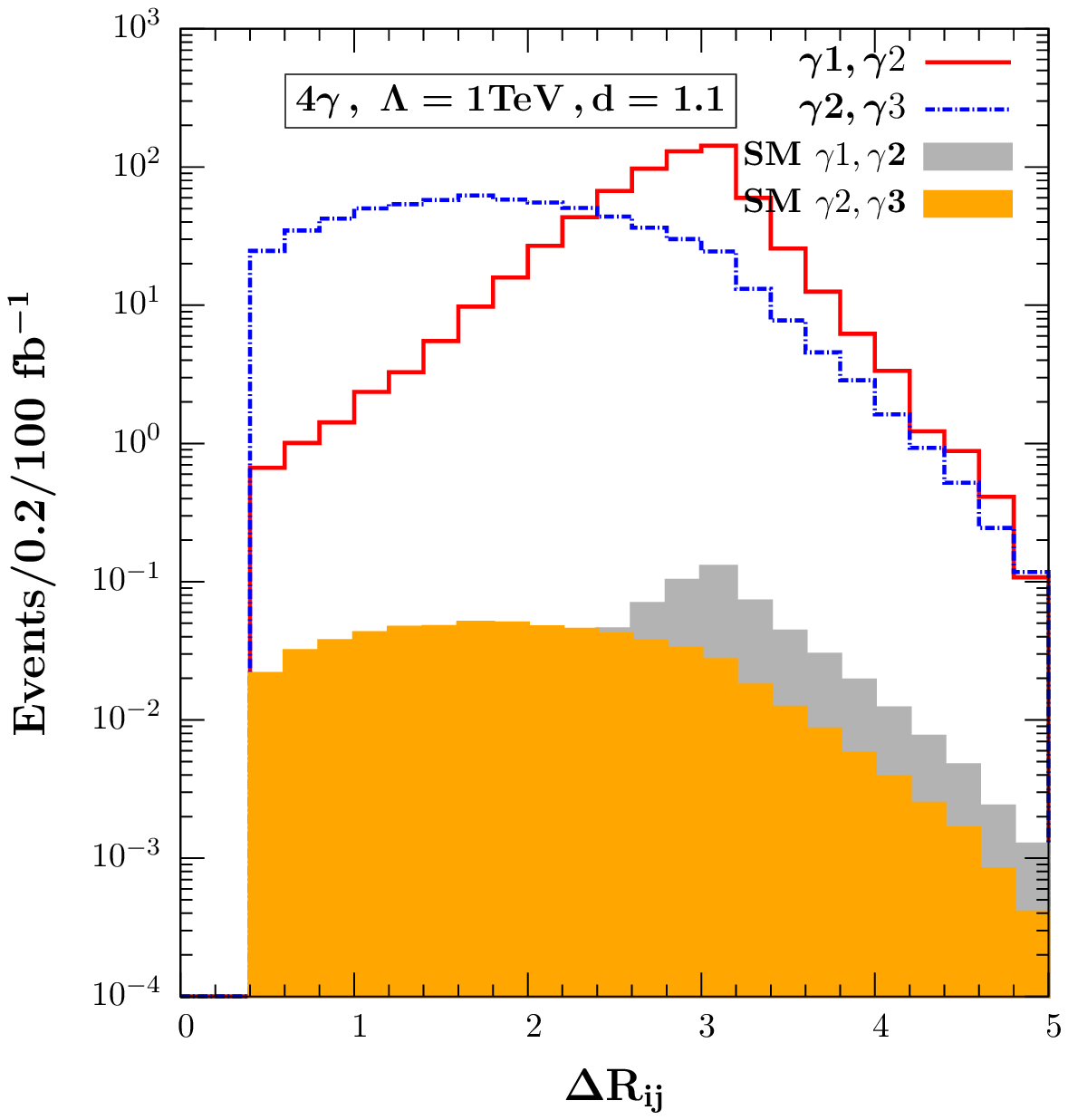} &\hspace*{-1cm}
				\includegraphics[width=3.5in,height=3.1in]{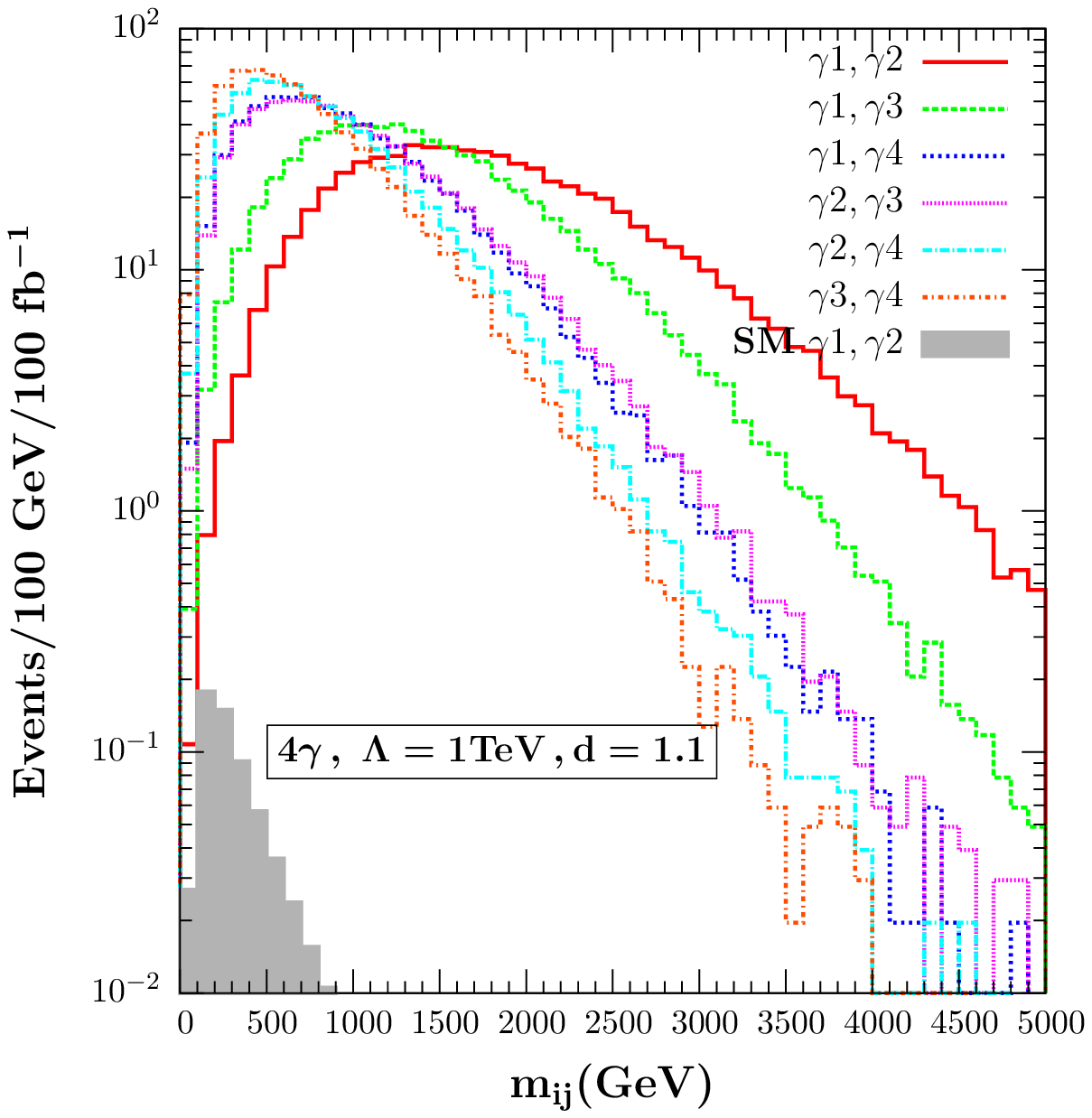}
				\end{array}$
			\end{center}
			\vskip -0.2in
			\caption{Various distributions for the $pp\to 4\gamma$ signal at LHC @ 14 TeV center of mass energy within a scalar unparticle scenario for different choices of $d$ and $\Lambda_{\cal U}$. In the case of invariant mass distribution, only the largest SM background is shown.  For the $\Delta R_{ij}$ distributions, two distinct SM backgrounds are preferred to be presented. $\Lambda_{\cal U}=3$ TeV case is not included in the $\Delta R_{ij}$ and $m_{ij}$ cases since it looks very much like the SM distribution.}\label{var4p}
		\end{figure}
			\begin{figure}[htb]
				\begin{center}
					\hspace*{-1.8cm}
					$\begin{array}{ccc}
					\includegraphics[width=2.6in,height=3.0in]{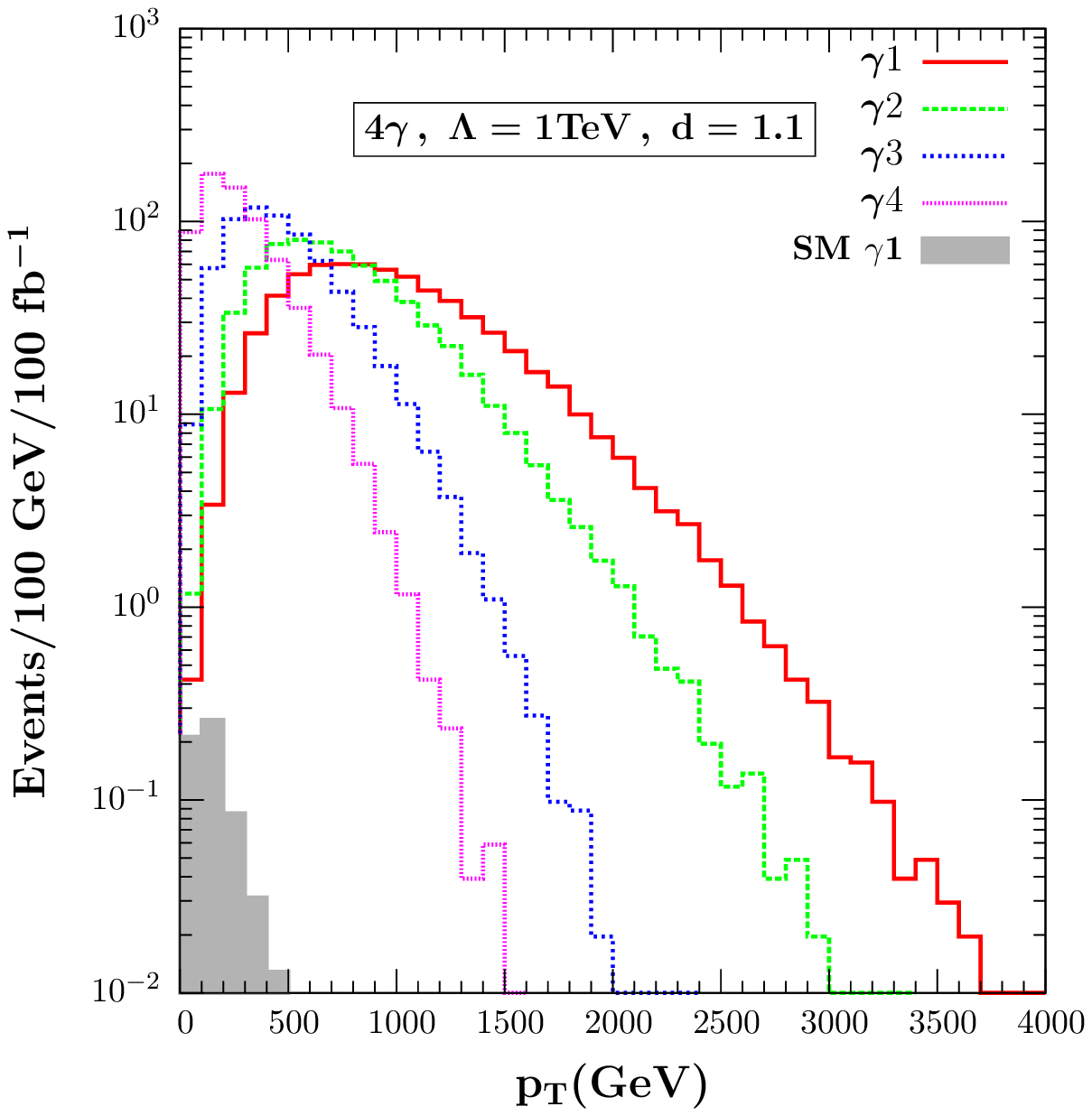} & \hspace*{-1cm}
					\includegraphics[width=2.6in,height=3.0in]{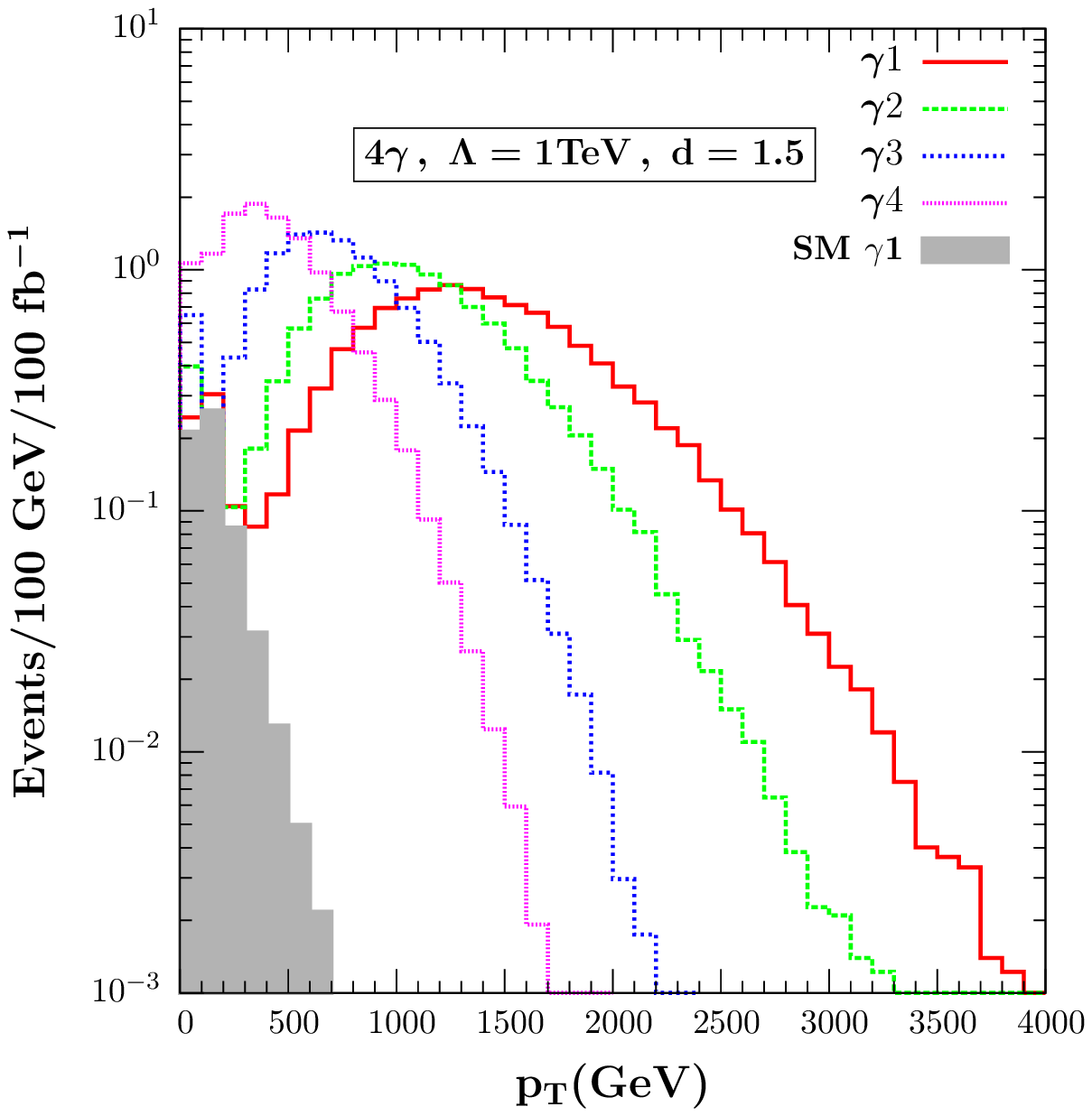} & \hspace*{-1cm} 
					\includegraphics[width=2.6in,height=3.0in]{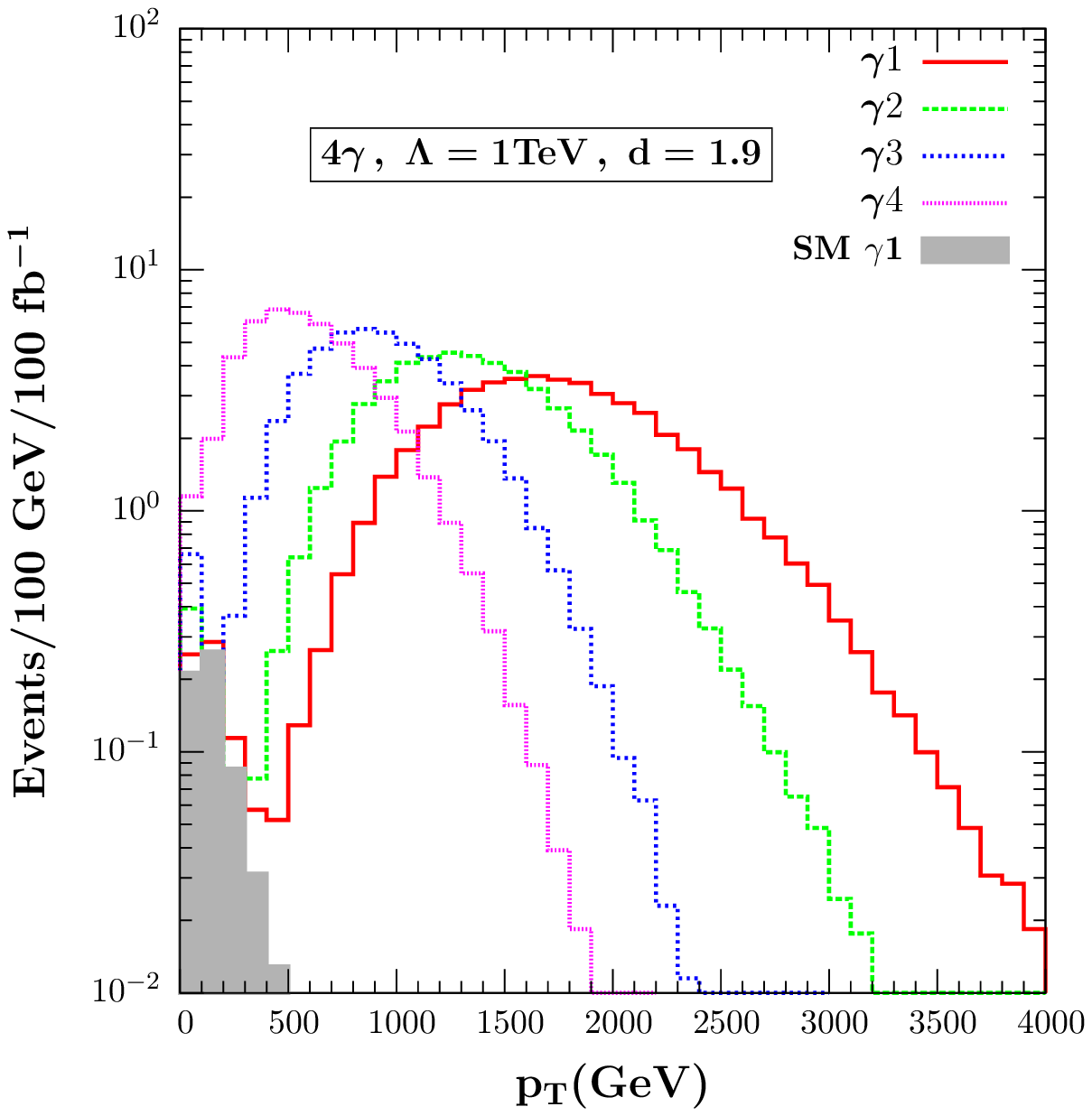} 
					\end{array}$
				\end{center}
				\vskip -0.2in
				\caption{The $p_T$ distributions of each photon at $\Lambda_{\cal U}=1$ TeV for $d=1.1,\,1.5,$ and $1.9$ within a scalar unparticle scenario. In each case, the largest SM background is depicted.}\label{pt4p}
			\end{figure}
			
	The number of events for the signal $pp \to \pppp$ at LHC with the center of mass energy $14$ TeV and the integrated luminosity $100\, {\rm fb}^{-1}$ are shown as a function of various variables in Fig.~\ref{var4p}. As far as the number of generated jets  is concerned the signal shows almost identical distributions with the largest SM case when $\Lambda_{\cal U}=3$ TeV but many more jets can be generated over the background for $\Lambda_{\cal U}=1$ TeV. For the case of $H_T$ distributions, the background shows a sharp drop and $H_T$ gets larger while the signal starts developing a shoulder for all cases with $\Lambda_{\cal U}=1$ TeV and for only $d_{\cal U}=1.1$ when $\Lambda_{\cal U}$ is taken 3 TeV. This practically means that heavy particles must be produced so that we get more events with large $H_T$. Additionally, an optimal $H_T$ cut value could be determined to reduce the background further if needed. The distributions with respect to the cone size for various photon pairs resemble each other (having similar peak patterns) when comparing the signal with the corresponding SM background. The number of background events are just subdued. As far as the topology of the events is concerned, among the hardest three photons, the distance between the hardest photon and the second hardest one peaks at larger values than the one between the second and the third. Hence, the hardest and the second hardest must come off from different branches. All possible invariant mass distributions are compared with the largest SM background and an invariant mass cut can further be fixed as well. The number of signal events as a function of the transverse momenta of the photons at a fixed $\Lambda_{\cal U}=1$ TeV for various $d$ values are presented in Fig.~\ref{pt4p}. In each case, only the largest SM background is included, and the photons are labeled in descending order based on their energies. It seems possible to eliminate the background altogether by using an improved cut value. In the case of $d_{\cal U}=1.5$ and $d_{\cal U}=1.9$ a higher luminosity might be needed for producing enough signal events.

	
	\subsection{$pp \to 2\gamma 2g$ Signal}

	\begin{figure}[htb]
	\begin{center}
\hspace*{-1.8cm}
	        $\begin{array}{cc}
		\includegraphics[width=3.5in,height=3.1in]{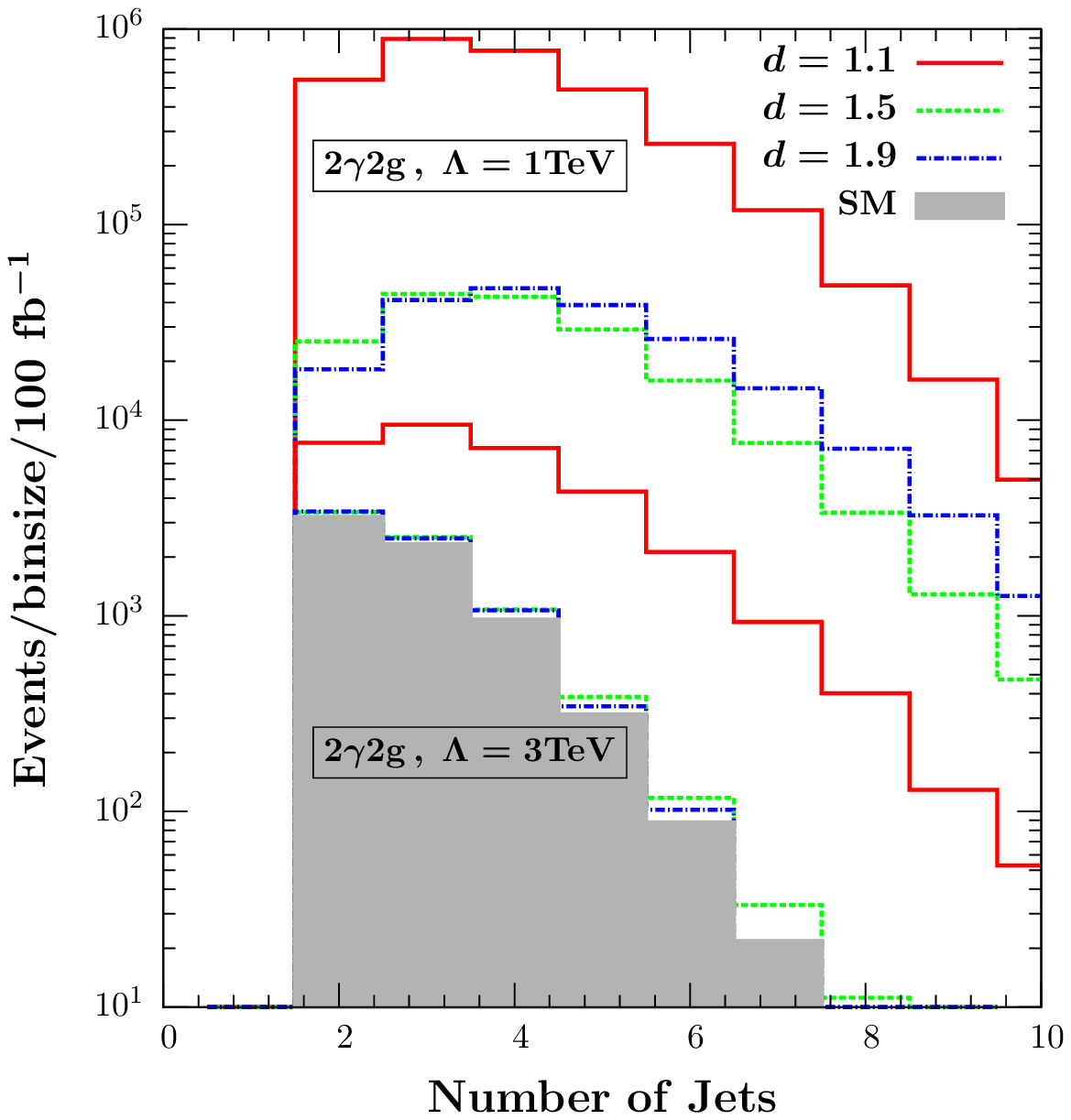} &\hspace*{-1cm}
		\includegraphics[width=3.5in,height=3.1in]{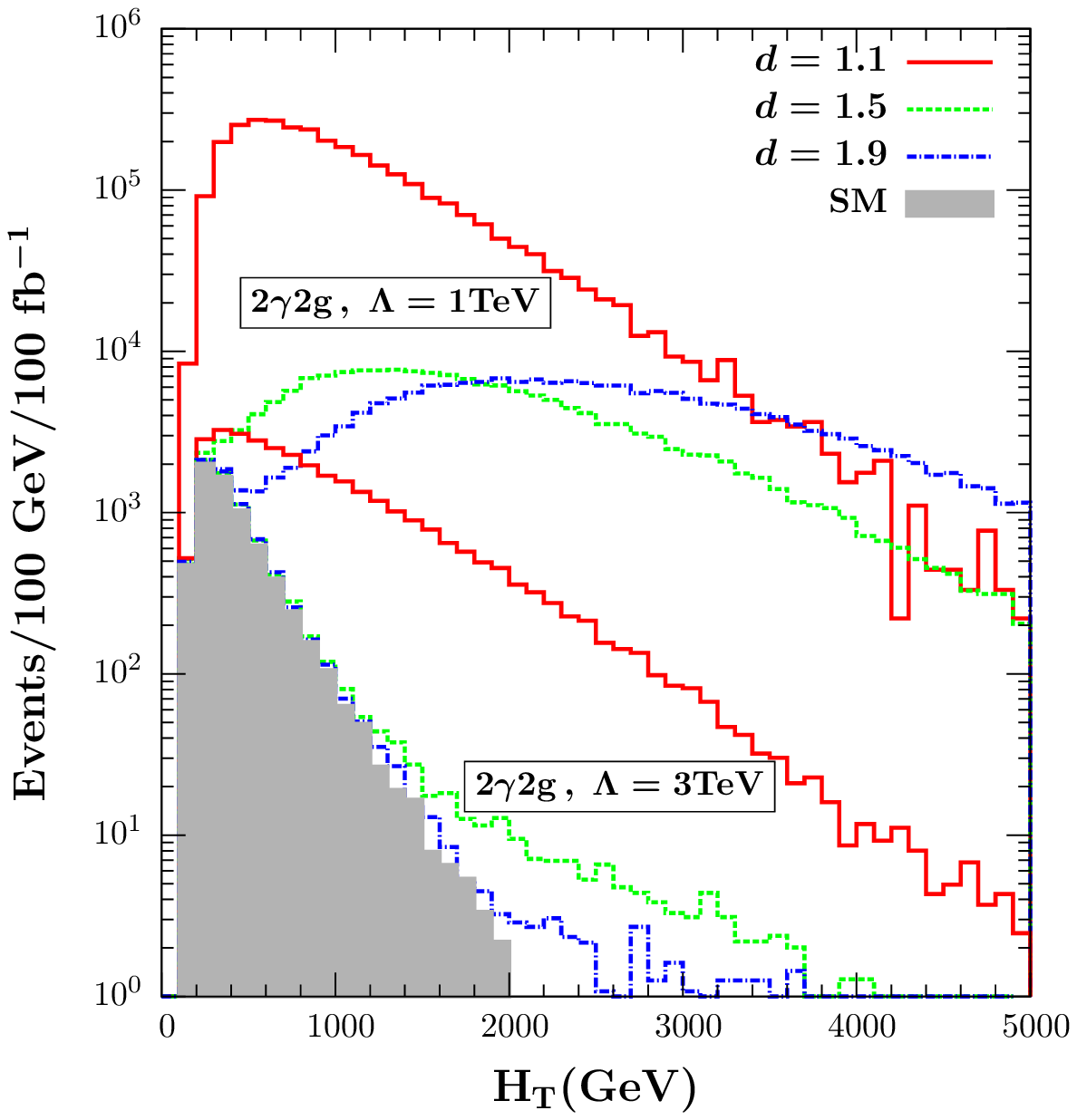} \\ \hspace{0.2cm}
		\includegraphics[width=3.4in,height=3.1in]{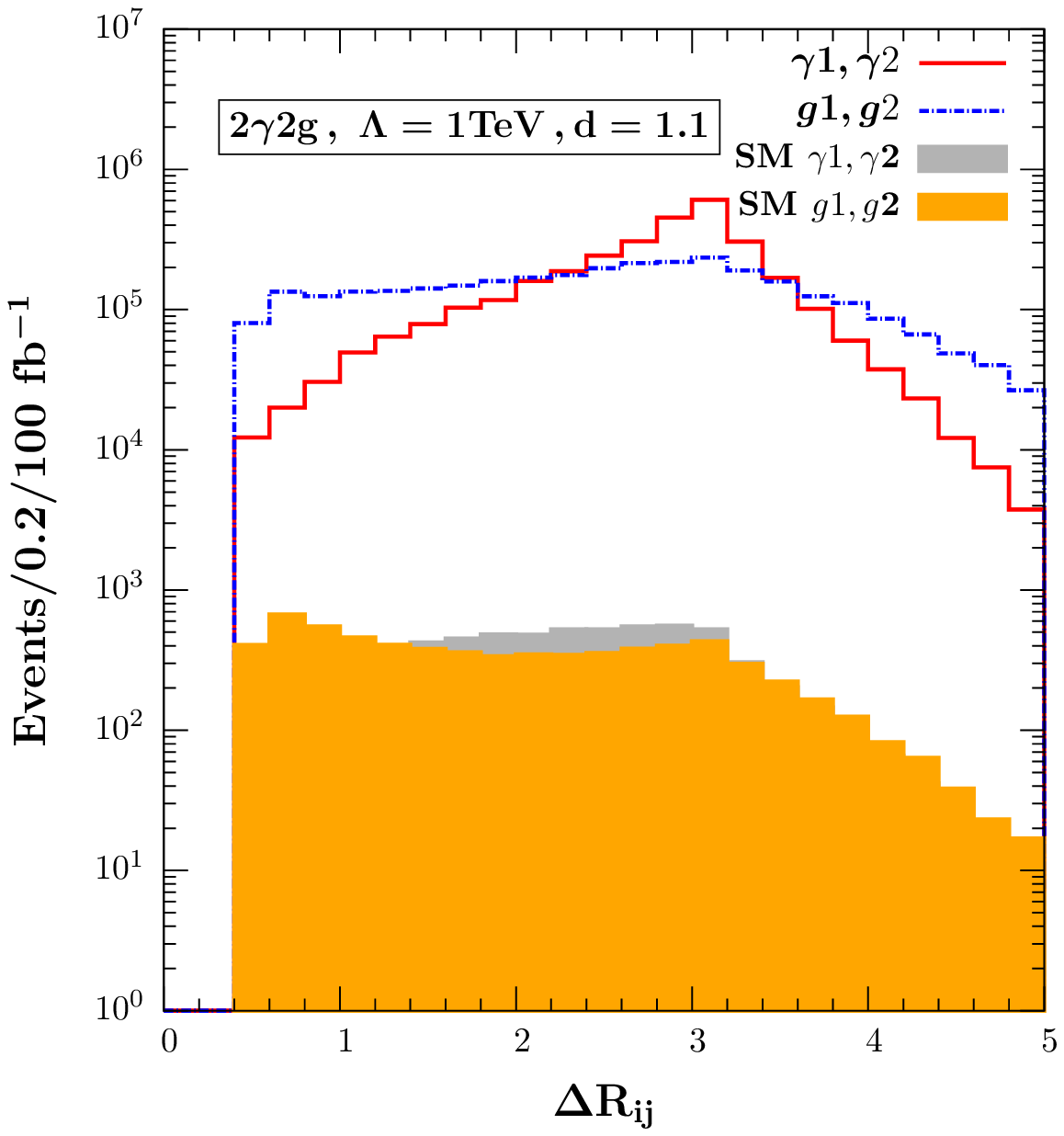} &\hspace*{-1cm}
		\includegraphics[width=3.5in,height=3.1in]{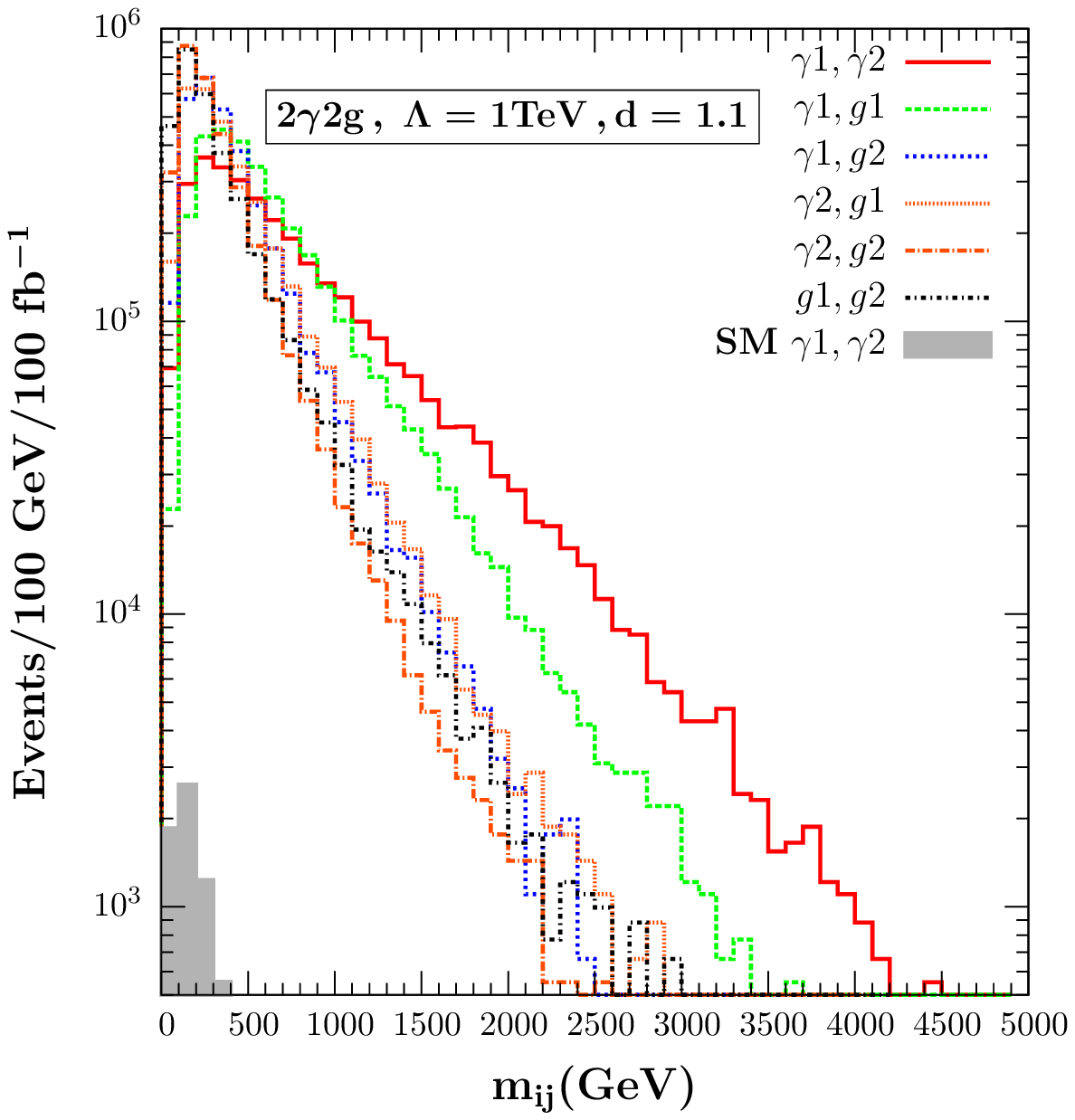}
	\end{array}$
	\end{center}
	\vskip -0.2in
	\caption{Various distributions for the $pp\to 2\gamma 2g$ signal within a scalar unparticle scenario for $\Lambda_{\cal U}=1$ TeV. In the case of invariant mass distribution, only the largest SM background is shown.  For the $\Delta R_{ij}$ distributions, two distinct SM backgrounds are preferred to be presented. $\Lambda_{\cal U}=3$ TeV case is not included in the $\Delta R_{ij}$ and $m_{ij}$ cases since it looks very much like the SM distribution.}\label{var2g2p}
	\end{figure}
	
	\begin{figure}[htb]
	\begin{center}
	\hspace*{-1.8cm}
	        $\begin{array}{ccc}
		\includegraphics[width=2.6in,height=3.1in]{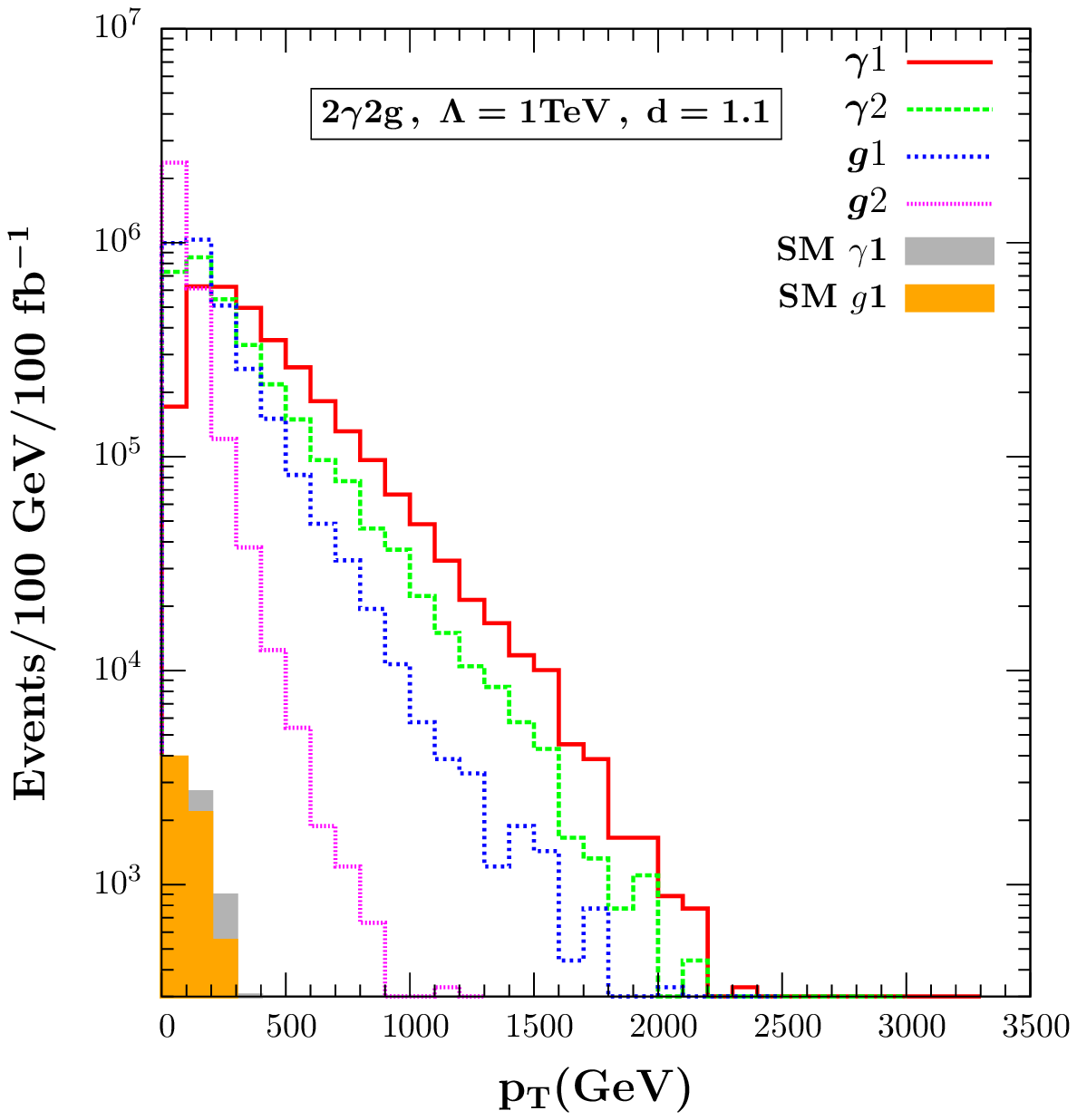} & \hspace*{-1cm}
		\includegraphics[width=2.6in,height=3.1in]{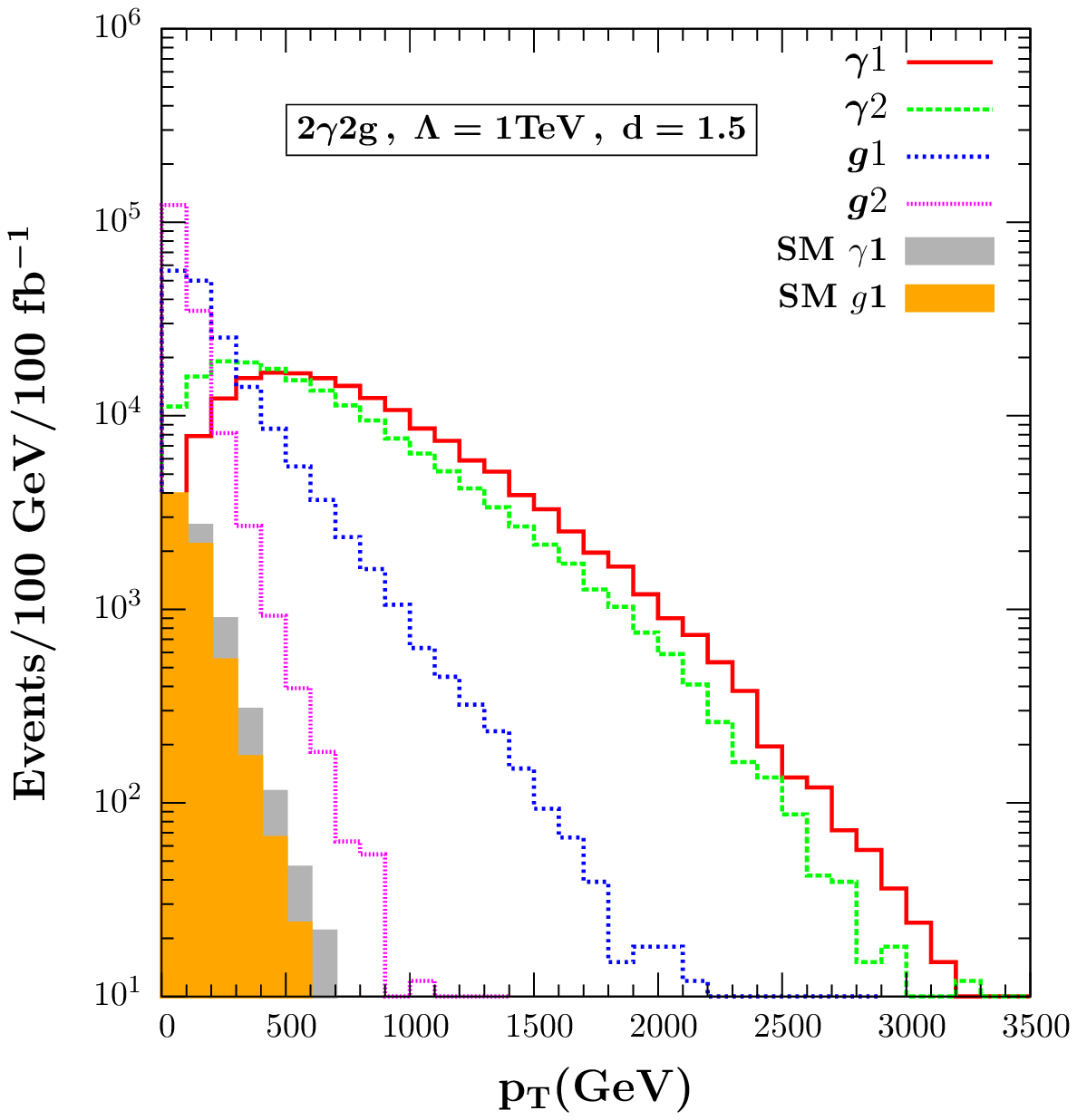} & \hspace*{-1cm} 
		\includegraphics[width=2.6in,height=3.1in]{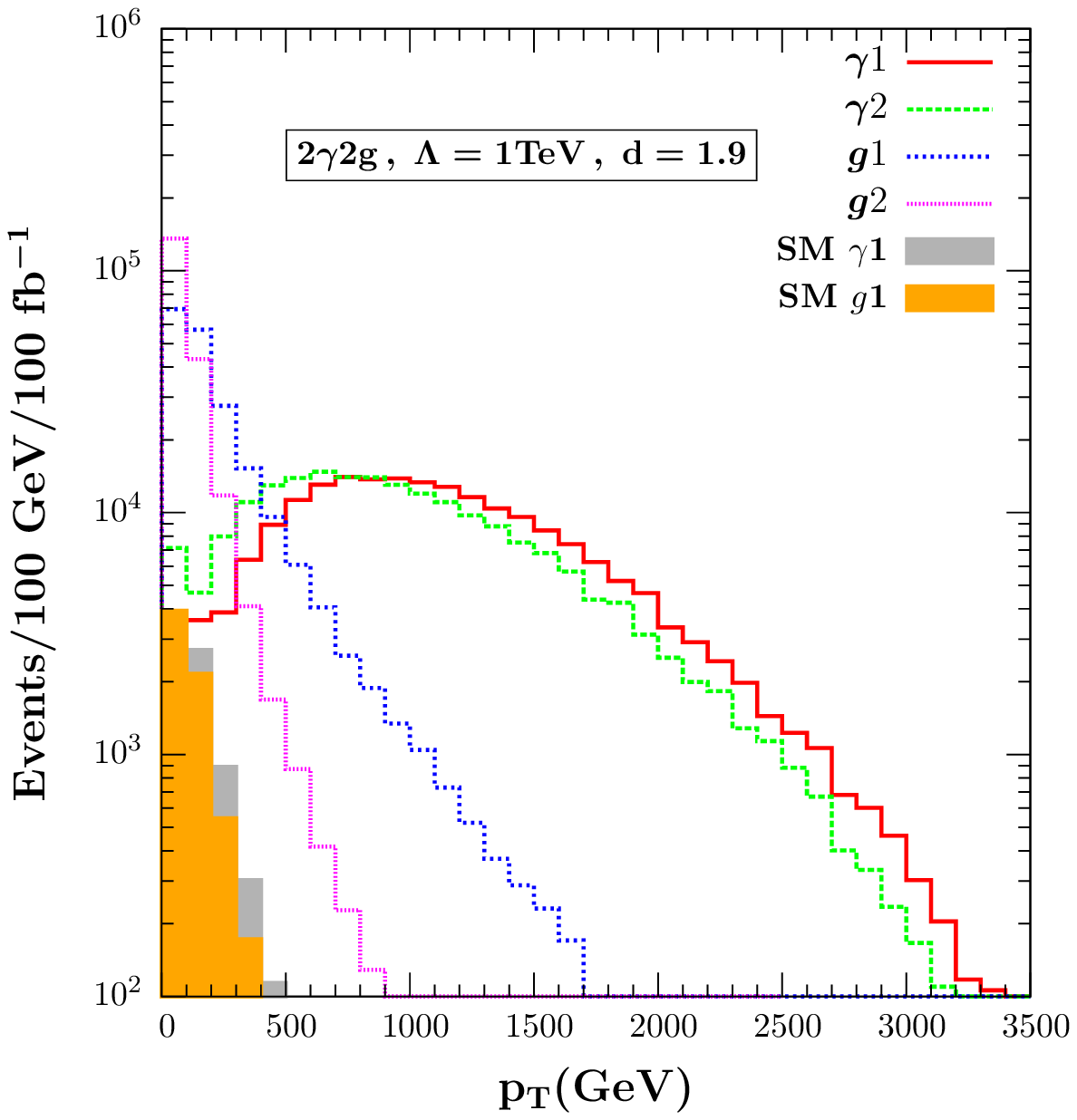} 
	\end{array}$
	\end{center}
	\vskip -0.2in
	\caption{The $p_T$ distributions of each photon at $\Lambda_{\cal U}=1$ TeV for $d_{\cal U}=1.1,\,1.5,$ and $1.9$ within a scalar unparticle scenario. In each case, the largest SM background is depicted.}\label{pt2g2p}
	\end{figure}
	
	As compared to the $4\gamma$ signal, here we require two photons and at least two gluon jets. We expect more events for both the signal and for the background. Our findings are depicted in Figs.~\ref{var2g2p} and \ref{pt2g2p}. The jet activity for the background is suppressed even at $\Lambda_{\cal U} = 3$ TeV for $d_{\cal U}=1.1$. Also from the $H_T$ distributions, we observe that the signal starts deviating from the background for even low energies at which the signal peaks. $\Delta R$ seems to be a useful quantity since the signal and background prefer to have peaks at opposite sites. For the transverse momentum distributions where the hardest photon and the hardest gluon jet distributions are included for the background, the background has almost no tail over 500 GeV while the signal shows much broader distributions with peaks moving to higher energies as $d_{\cal U}$ gets larger. 
	

	\subsection{$pp \to 2\gamma 2l $ Signal}

	\begin{figure}[b]
	\begin{center}
\hspace*{-1.8cm}
	        $\begin{array}{cc}
		\includegraphics[width=3.5in,height=3.1in]{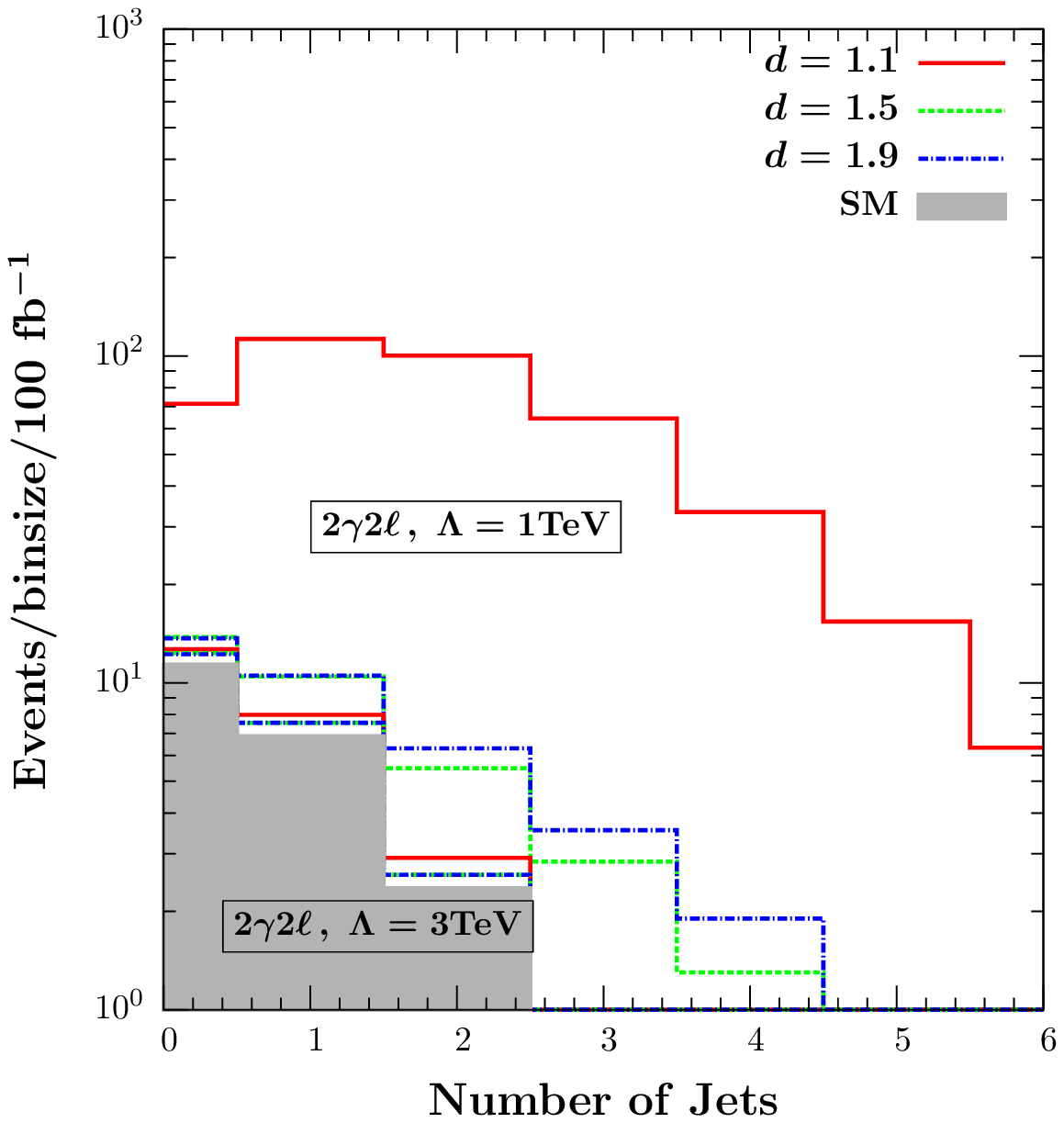} &\hspace*{-1cm}
		\includegraphics[width=3.5in,height=3.1in]{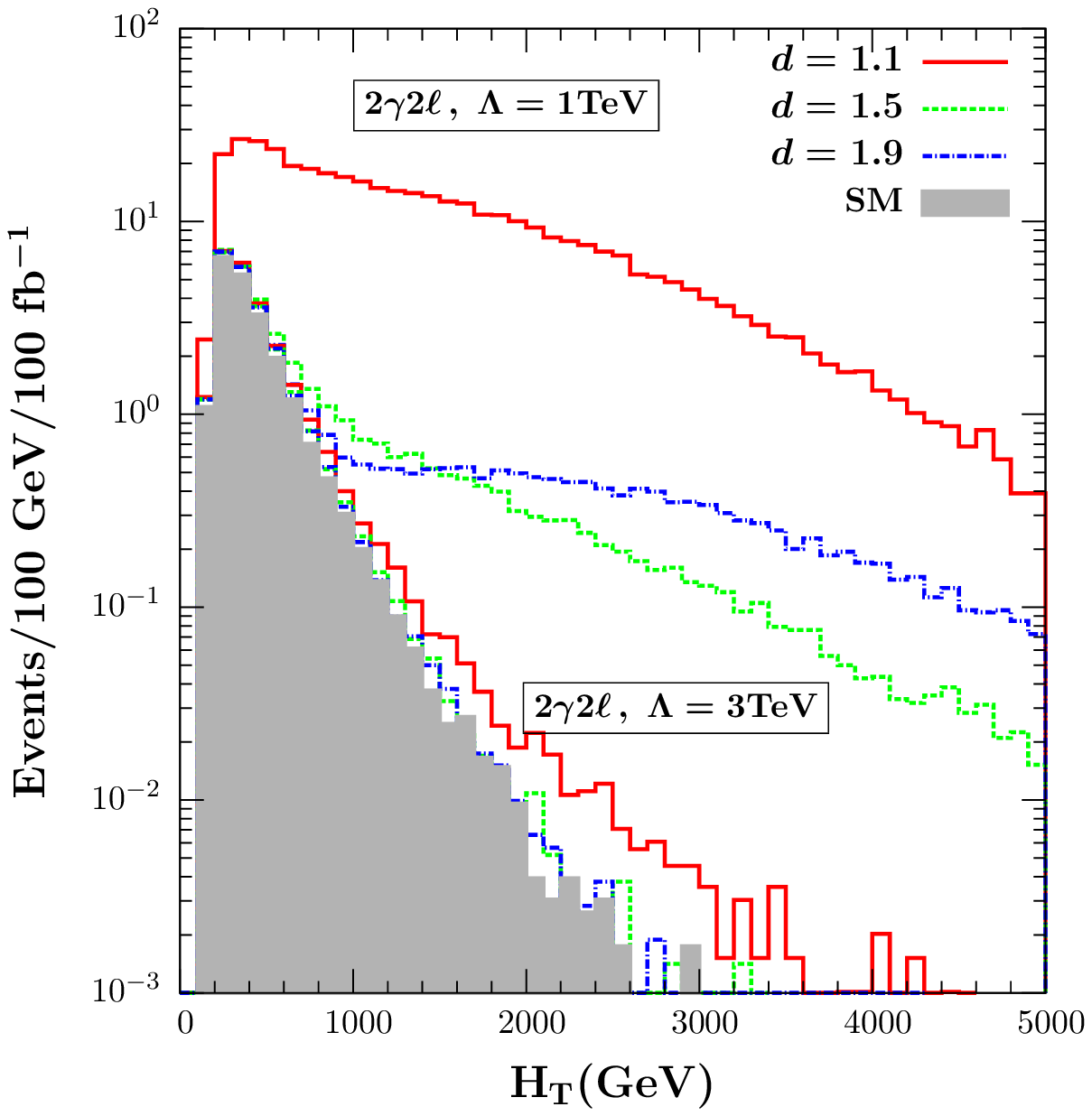} \\ \hspace{0.2cm}
		\includegraphics[width=3.4in,height=3.1in]{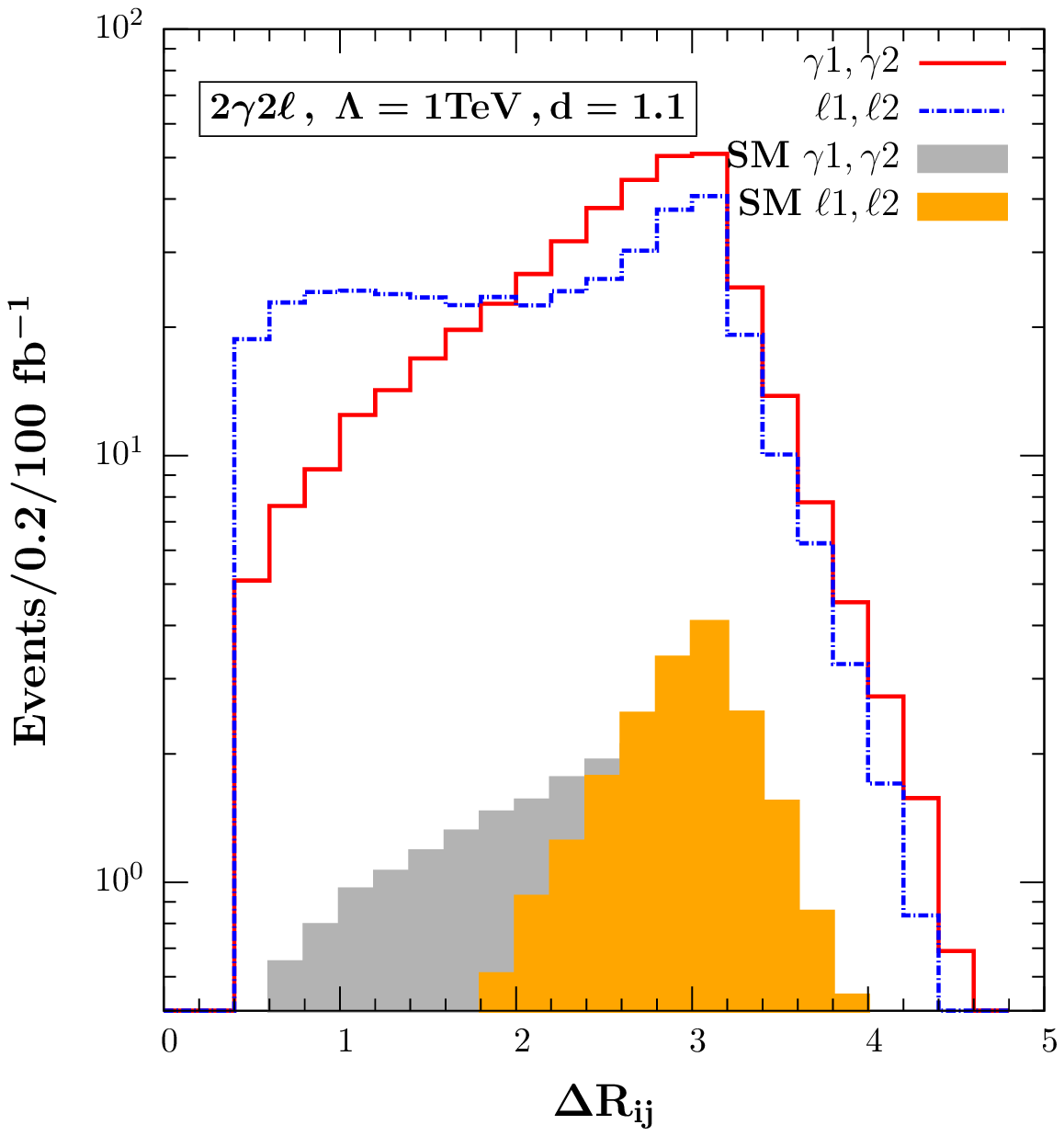} &\hspace*{-1cm}
		\includegraphics[width=3.5in,height=3.1in]{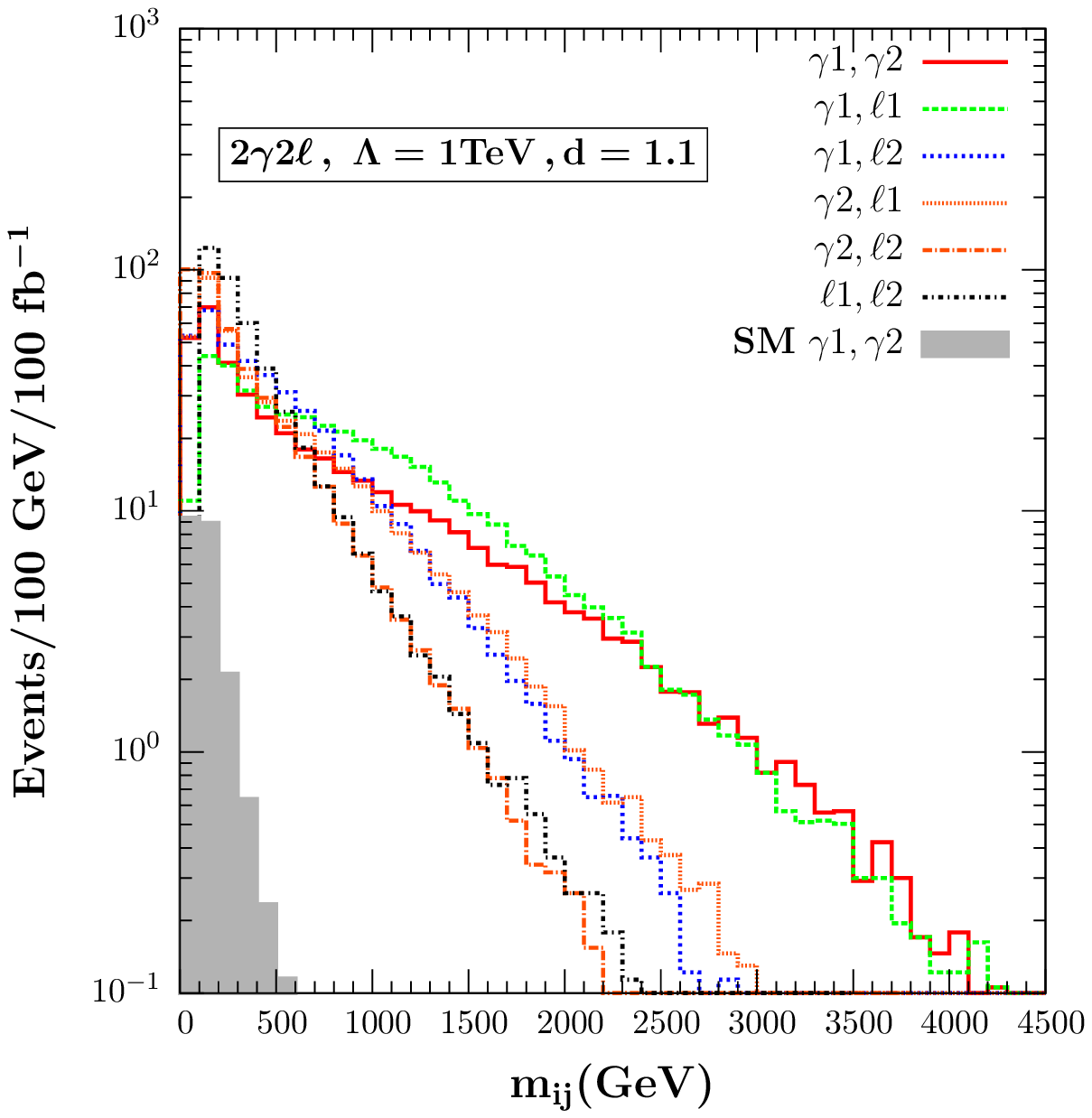}
	\end{array}$
	\end{center}
	\vskip -0.2in
	\caption{Various distributions for the $pp\to 2\gamma 2\ell$ signal within a scalar unparticle scenario for $\Lambda_{\cal U}=1$ TeV. In the case of invariant mass distribution, only the largest SM background is shown.  For the $\Delta R_{ij}$ distributions, two distinct SM backgrounds are preferred to be presented. $\Lambda_{\cal U}=3$ TeV case is not included in the $\Delta R_{ij}$ and $m_{ij}$ cases since it looks very much like the SM distribution.}\label{var2p2l}
	\end{figure}
	
	\begin{figure}[htb]
	\begin{center}
	\hspace*{-1.8cm}
	        $\begin{array}{ccc}
		\includegraphics[width=2.6in,height=3.1in]{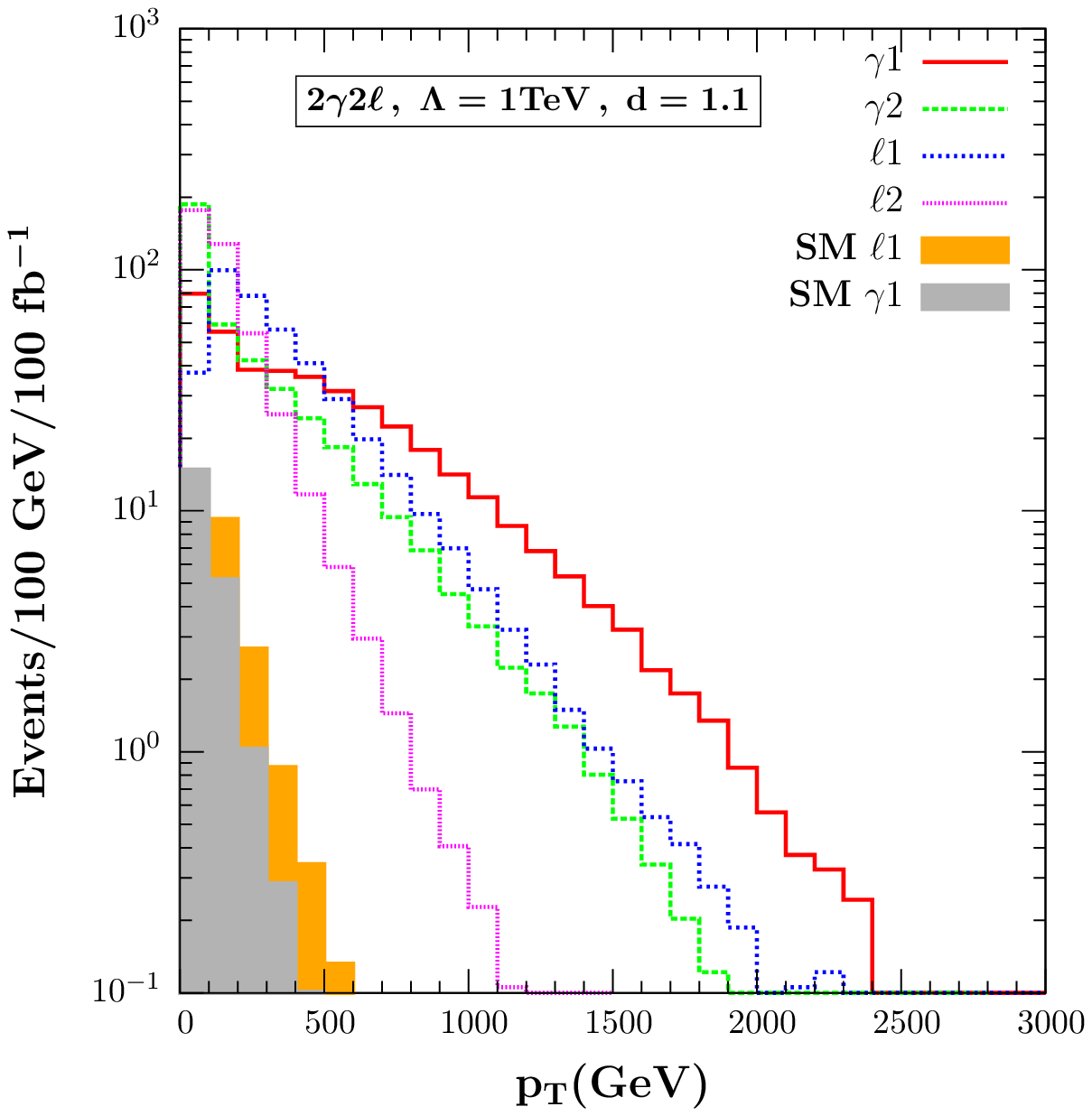} & \hspace*{-1cm}
		\includegraphics[width=2.6in,height=3.1in]{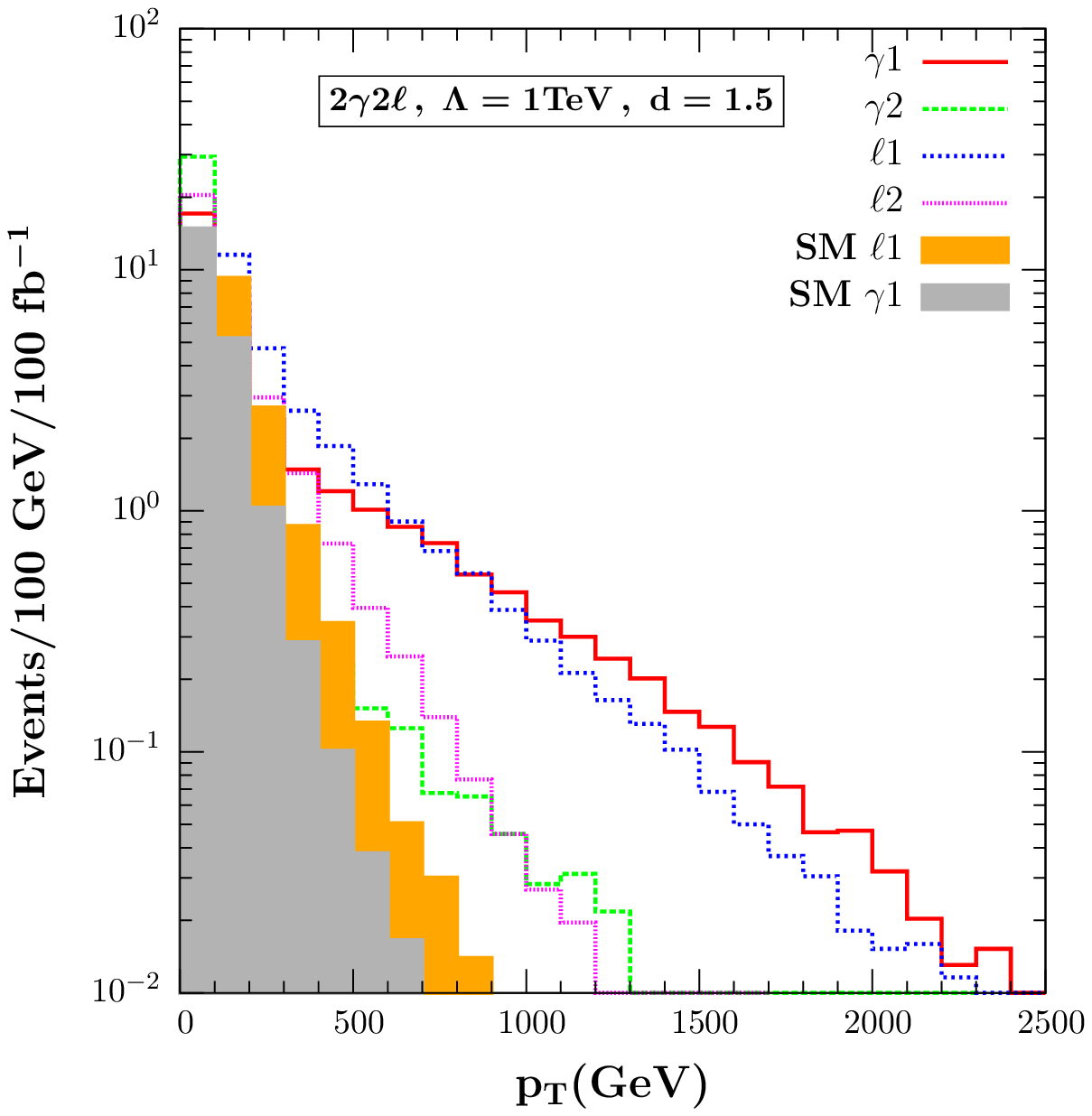} & \hspace*{-1cm} 
		\includegraphics[width=2.6in,height=3.1in]{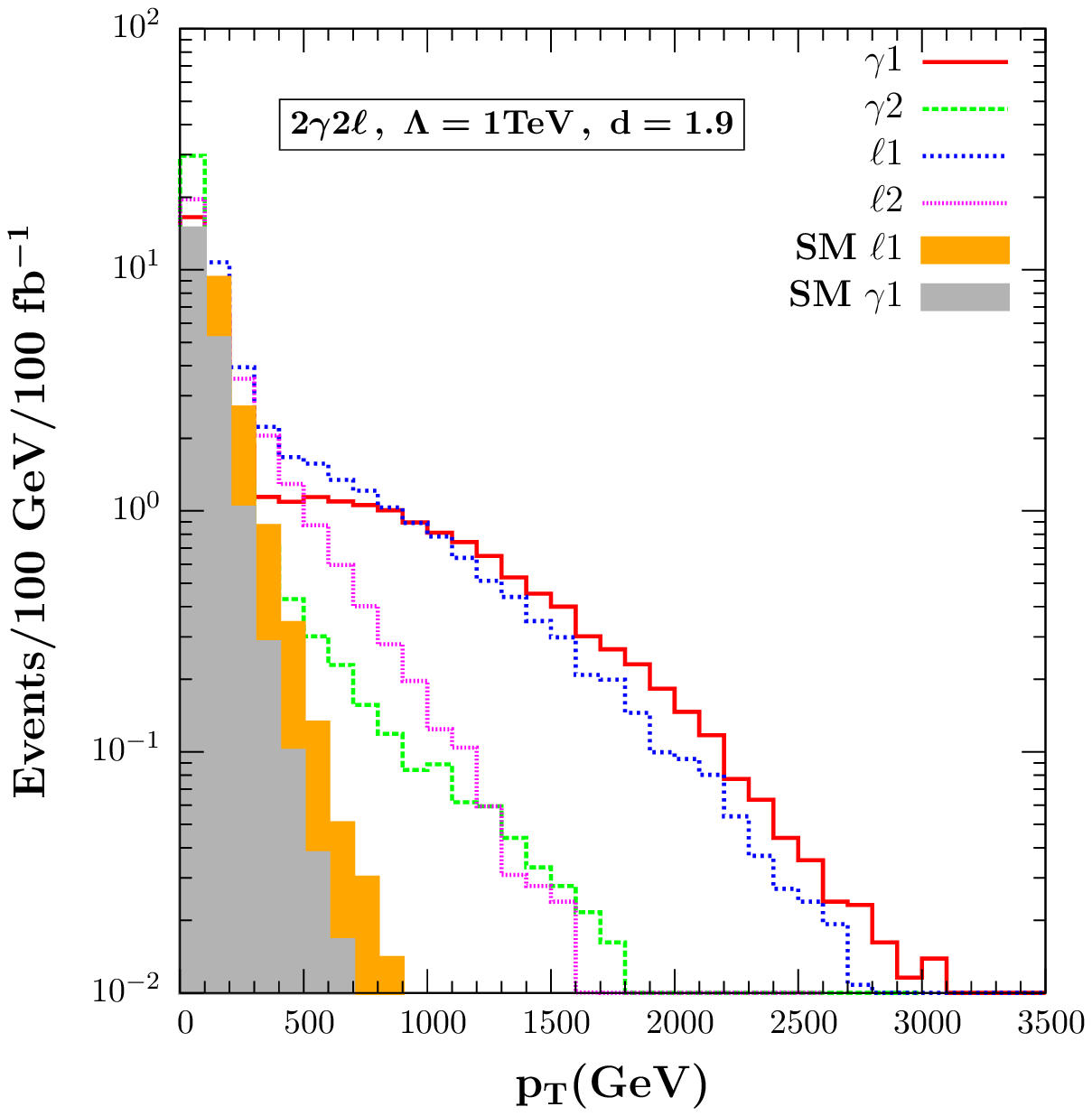} 
	\end{array}$
	\end{center}
	\vskip -0.2in
	\caption{The $p_T$ distributions of each photon at $\Lambda_{\cal U}=1$ TeV for $d_{\cal U}=1.1,\,1.5,$ and $1.9$ within a scalar unparticle scenario. In each case, the largest SM background is depicted.}\label{pt2p2l}
	\end{figure}
If we consider the signal with two photons and two isolated charged leptons where leptons could be electrons or muons or both, the results are summarized in Figs. \ref{var2p2l} and \ref{pt2p2l}. The jet activities for the signal and background resemble each other except for $\Lambda_{\cal U}=1$ TeV and $d_{\cal U}=1.1$. $H_T$ distribution shows that the signal shrinks to the background for $\Lambda_{\cal U}=3$ TeV and above. For $\Lambda_{\cal U}=1$ TeV, the signal starts dominating the background around the energy scale 1 TeV. Both the invariant mass and the transverse momenta distributions have similar features, showing enhancement especially at high energy tail.

	\subsection{$pp \to 2e2\mu$ Signal}
	
		\begin{figure}[htb]
	\begin{center}
\hspace*{-1.8cm}
	        $\begin{array}{cc}
		\includegraphics[width=3.5in,height=3.1in]{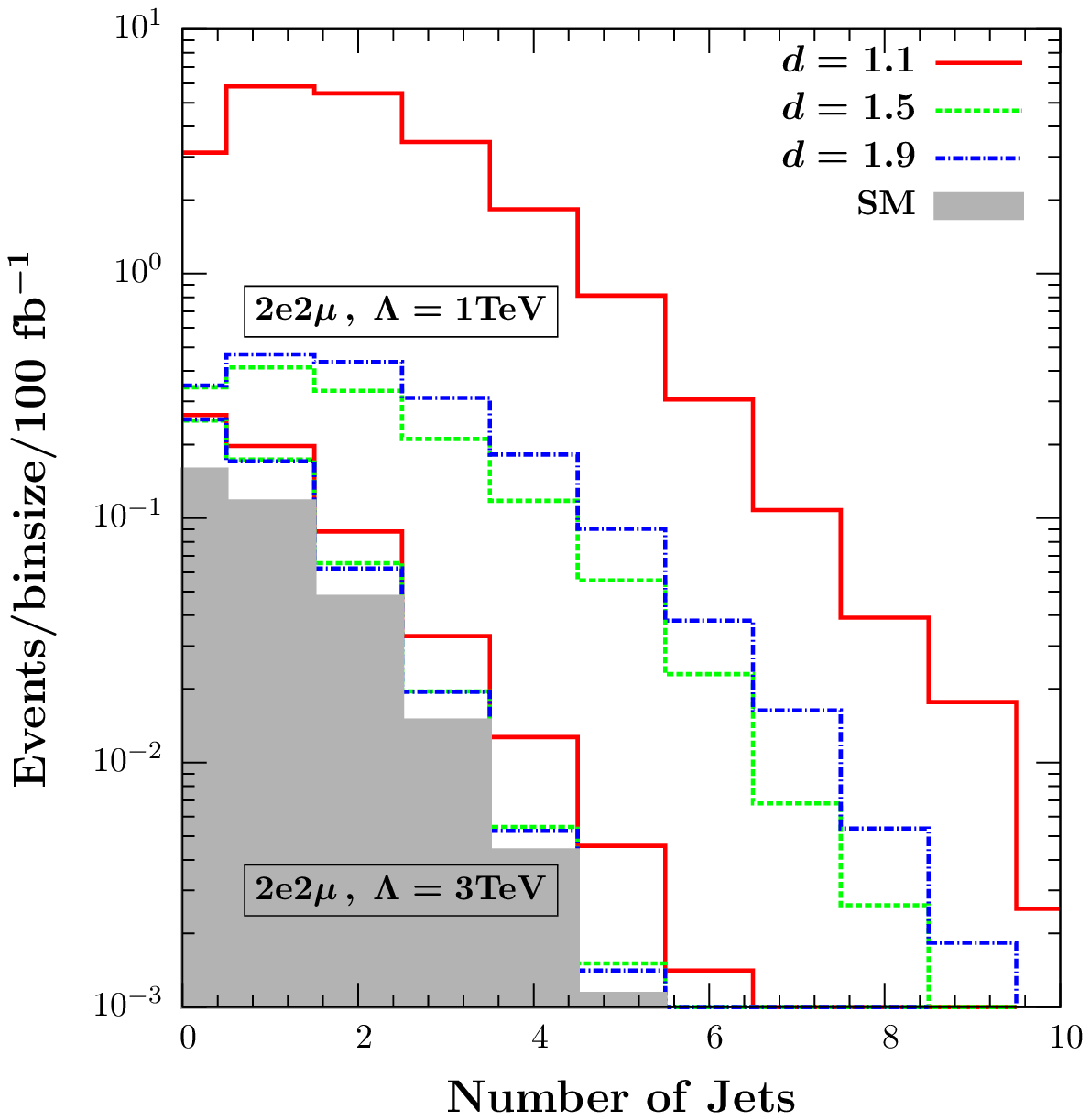} &\hspace*{-1cm}
		\includegraphics[width=3.5in,height=3.1in]{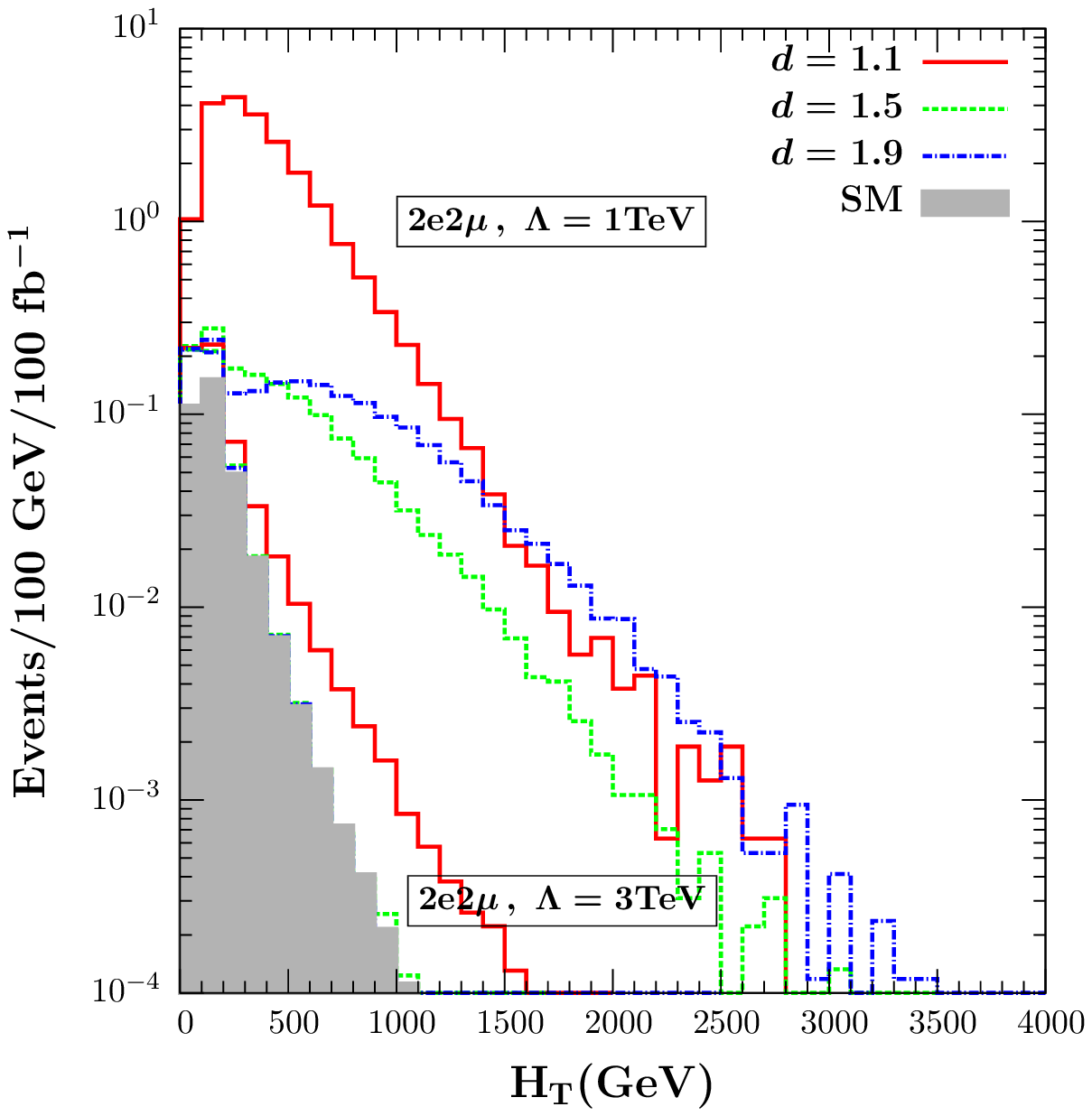} \\ \hspace{0.2cm}
		\includegraphics[width=3.4in,height=3.1in]{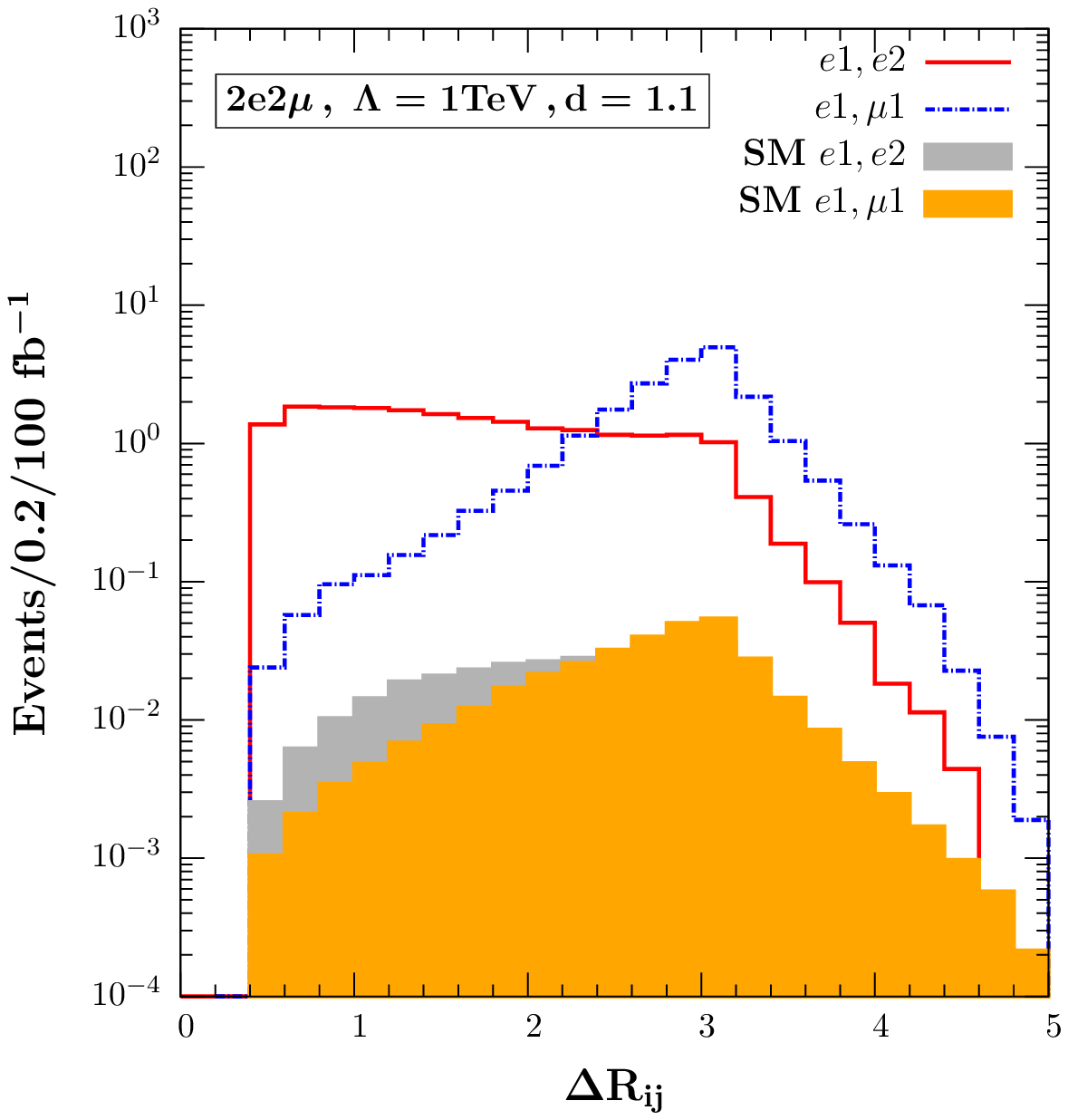} &\hspace*{-1cm}
		\includegraphics[width=3.5in,height=3.1in]{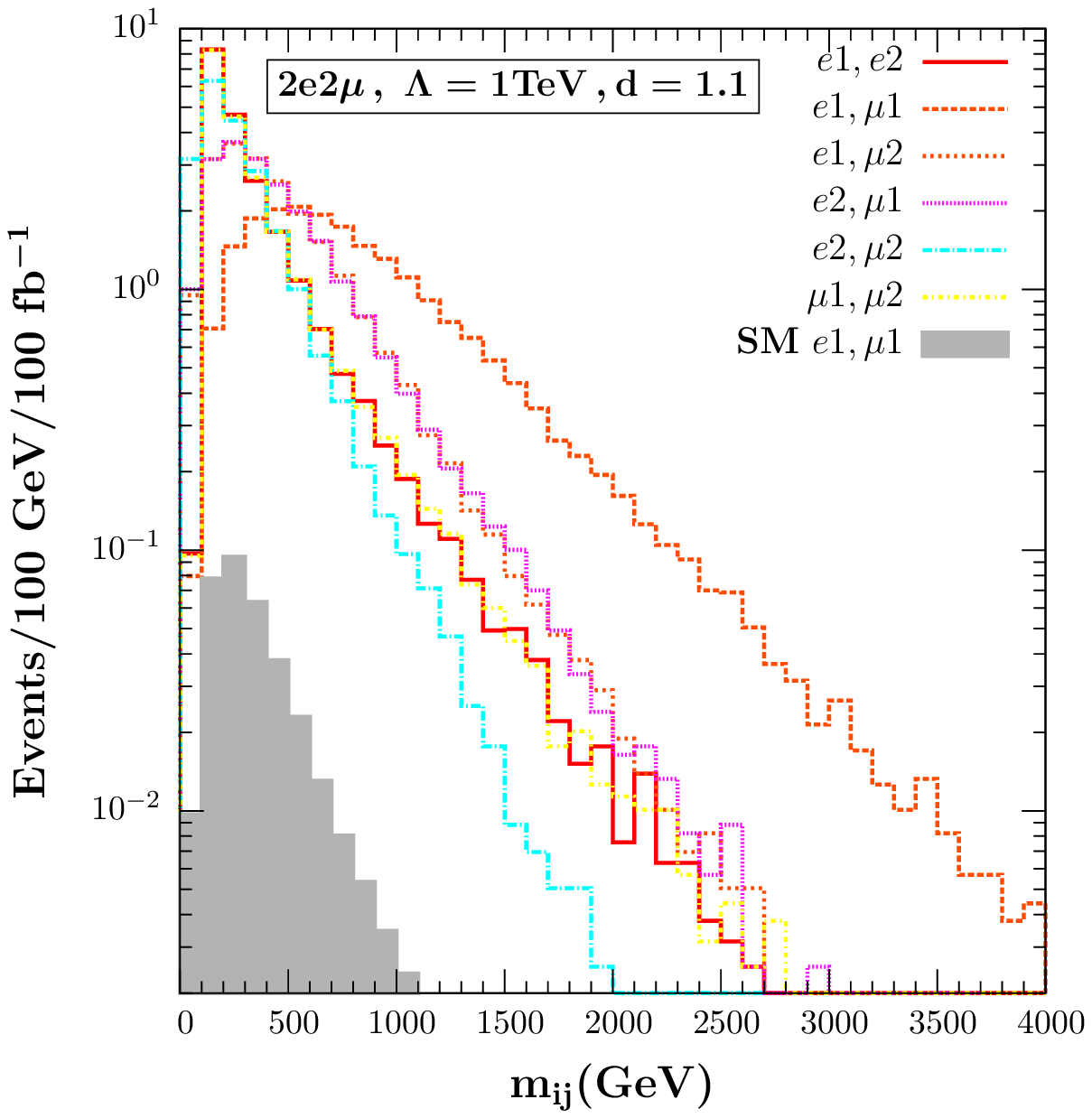}
	\end{array}$
	\end{center}
	\vskip -0.2in
	\caption{Various distributions for the $pp\to 2e 2\mu$ signal within a scalar unparticle scenario for $\Lambda_{\cal U}=1$ TeV. In the case of invariant mass distribution, only the largest SM background is shown.  For the $\Delta R_{ij}$ distributions, two distinct SM backgrounds are preferred to be presented. $\Lambda_{\cal U}=3$ TeV case is not included in the $\Delta R_{ij}$ and $m_{ij}$ cases since it looks very much like the SM distribution.}\label{var2e2m}
	\end{figure}

	\begin{figure}[h!]
	\begin{center}
	\hspace*{-1.8cm}
	        $\begin{array}{ccc}
		\includegraphics[width=2.6in,height=3.1in]{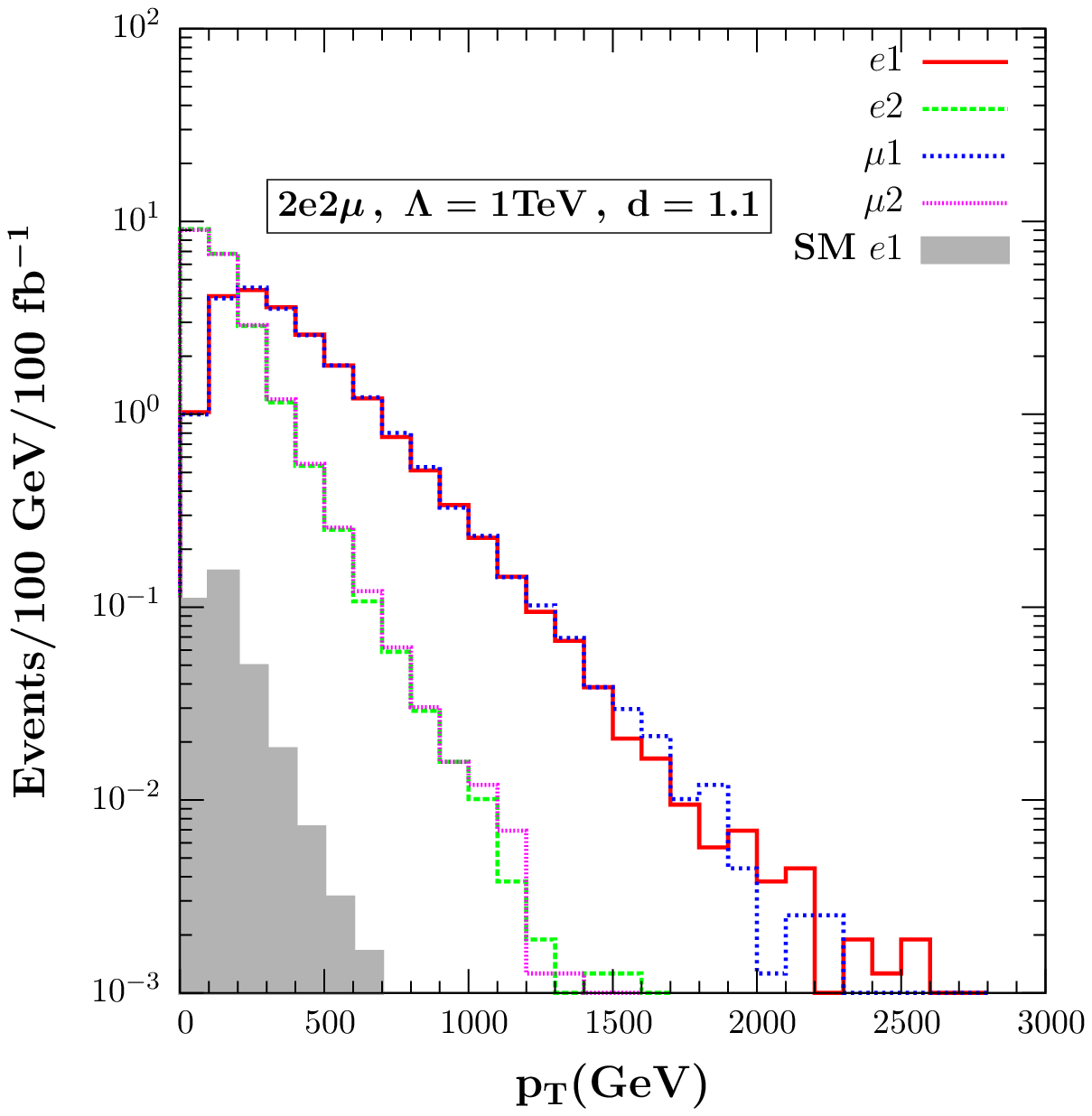} & \hspace*{-1cm}
		\includegraphics[width=2.6in,height=3.1in]{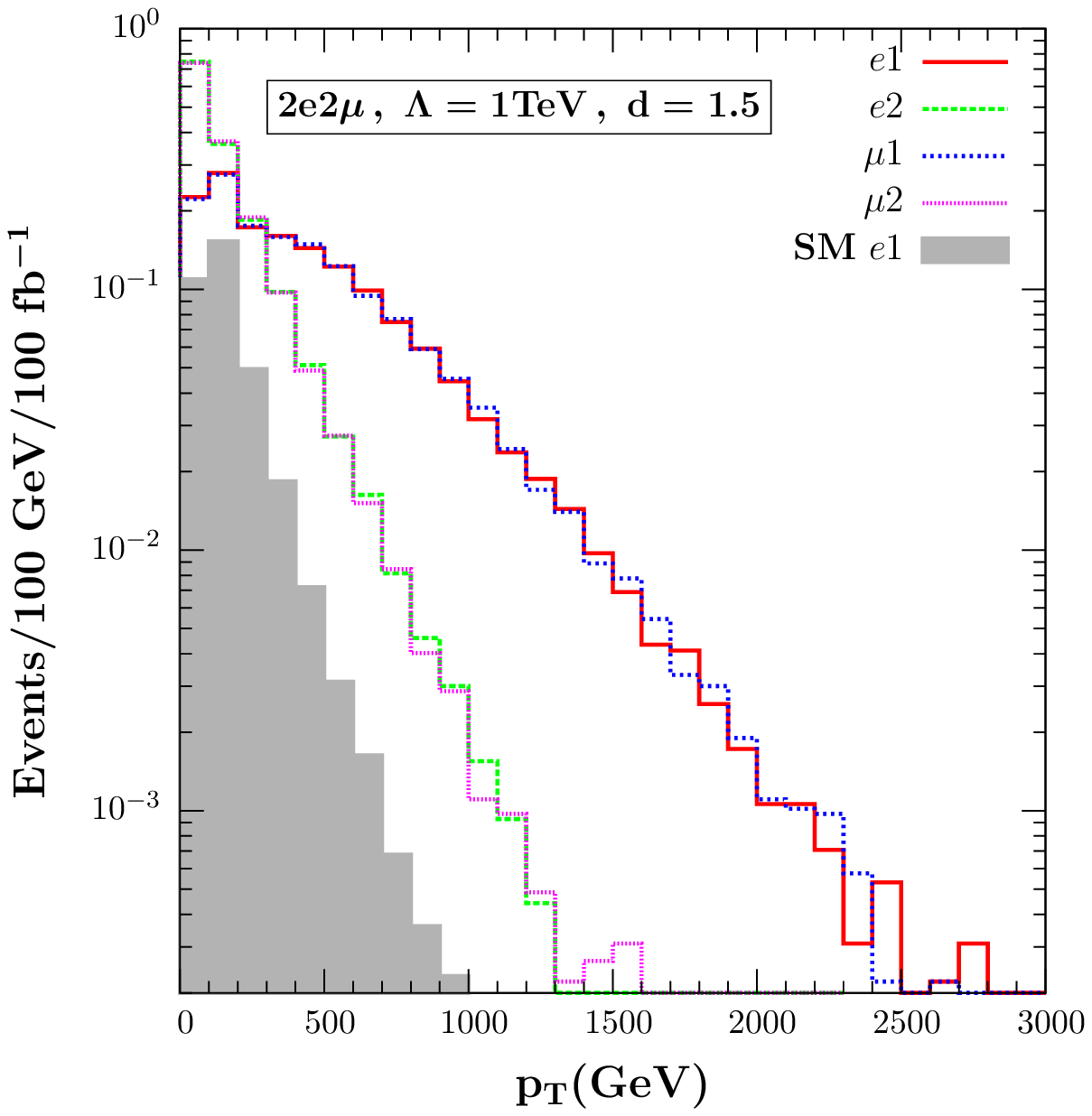} & \hspace*{-1cm} 
		\includegraphics[width=2.6in,height=3.1in]{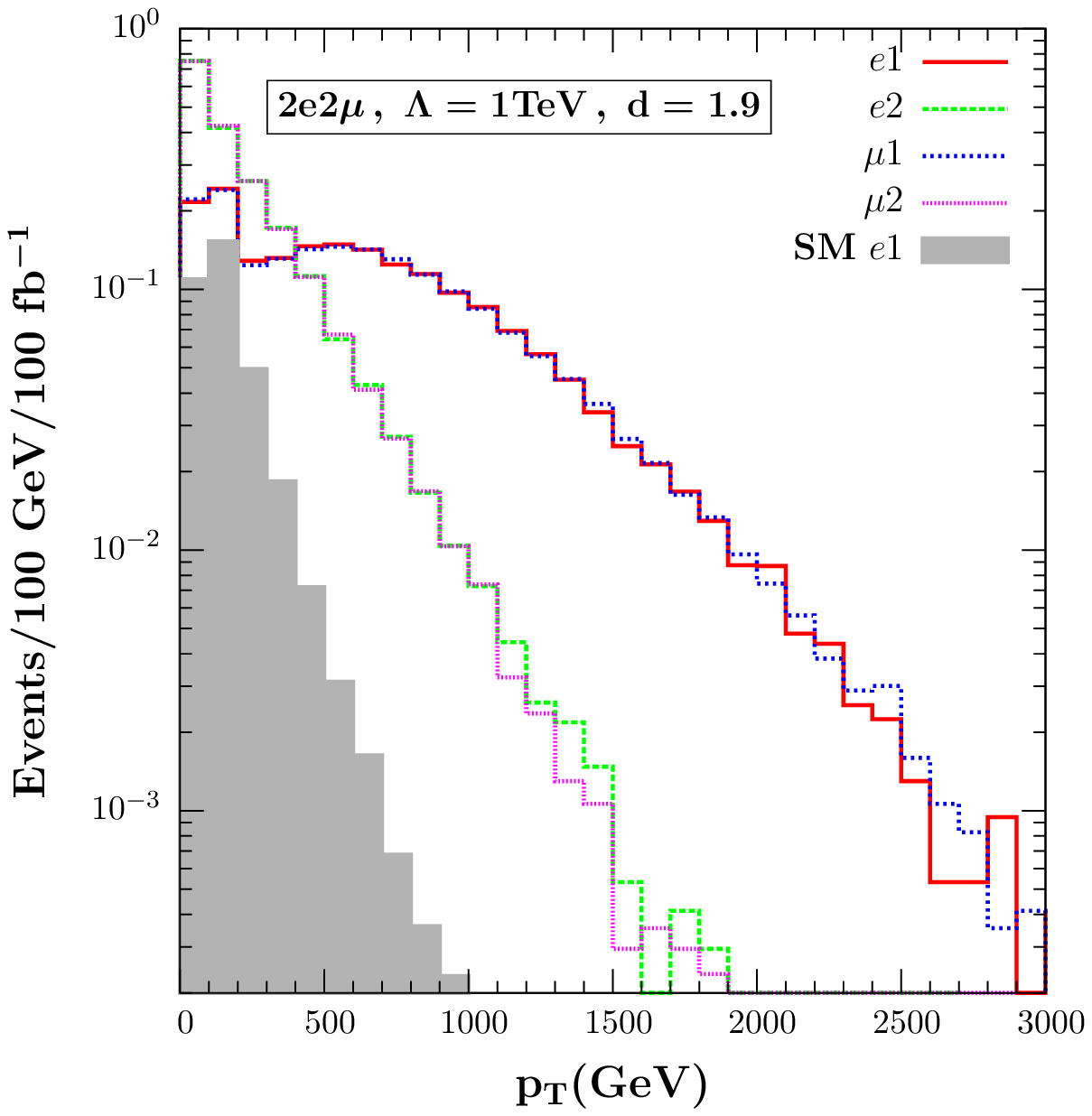} 
	\end{array}$
	\end{center}
	\vskip -0.2in
	\caption{The $p_T$ distributions of each photon at $\Lambda_{\cal U}=1$ TeV for $d=1.1,\,1.5,$ and $1.9$ within a scalar unparticle scenario. In each case, the largest SM background is depicted.}\label{pt2e2m}
	\end{figure}
\def\arraystretch{0.8}
\begin{table}[h!]
	\caption{The summary of the numerical analysis of all signals for various $(d_{\cal U},\Lambda_{\cal U})$ values including cross sections, expected signal events as well as the significance defined as $S/\sqrt{S+B}$ where $B$ stands for the background events. The center of mass energy is $14$ TeV with integrated luminosity 100$fb^{-1}$.}
	\label{tab:significance}
	\begin{tabular}{l@{\hskip 0.4in}c@{\hskip 0.4in}c@{\hskip 0.4in}c@{\hskip 0.4in}c@{\hskip 0.4in}c}
		\hline
		\hline
		Process & $\Lambda_{\cal U}$ & $d_{\cal U}$ & $\sigma(pb)$ & $S$ & $\frac{S}{\sqrt{S+B}}$ \\ \midrule
		\multirow{6}{*}{$pp \rightarrow 4 \gamma$} & \multirow{3}{*}{$1~TeV$}  & $1.1$ & $9.792 \times 10^{-3}$ & 949.99 $\pm$ 5.32 & 30.8095 $\pm$ 0.0865 \\  
		& & $1.5$ & $1.745 \times 10^{-4}$ & 16.890 $\pm$ 0.736 & 4.019 $\pm$ 0.098 \\
		& & $1.9$ & $7.665 \times 10^{-4}$ & 74.48 $\pm$ 1.45 & 8.5860 $\pm$ 0.0864 \\ \cmidrule{2-6}
		& \multirow{3}{*}{$3~TeV$}  & $1.1$ & $1.077 \times 10^{-5}$ & 0.95 $\pm$ 0.34 & 0.722 $\pm$ 0.199 \\
		& & $1.5$ & $1.018 \times 10^{-5}$  & 0.889 $\pm$ 0.336 & 0.691 $\pm$ 0.202 \\
		& & $1.9$ & $1.017 \times 10^{-5}$ & 0.889 $\pm$ 0.335 & 0.691 $\pm$ 0.201 \\ \midrule
		\multirow{6}{*}{$pp \rightarrow 2\gamma 2g$} & \multirow{3}{*}{$1~TeV$}  & $1.1$ & $5.520 \times 10^{1}$ & 1579879 $\pm$ 1061 & 1254.288 $\pm$ 0.424 \\
		& & $1.5$ & $3.010 \times 10^{0}$ & 80926 $\pm$ 243 & 273.432 $\pm$ 0.453 \\
		& & $1.9$ & $3.798 \times 10^{0}$ & 82830 $\pm$ 254 & 276.872 $\pm$ 0.467 \\ \cmidrule{2-6}
		& \multirow{3}{*}{$3~TeV$}  & $1.1$ & $6.166 \times 10^{-1}$ & 19636 $\pm$ 115 & 121.07 $\pm$ 0.47 \\
		& & $1.5$ & $1.826 \times 10^{-1}$ & 7143.3 $\pm$ 65.9 & 60.780 $\pm$ 0.439 \\
		& & $1.9$ & $1.797 \times 10^{-1}$ & 7121.5 $\pm$ 65.6 & 60.642 $\pm$ 0.437 \\ \midrule

		\multirow{6}{*}{$pp \rightarrow 2\gamma 2l$} & \multirow{3}{*}{$1~TeV$}  & $1.1$ & $8.117 \times 10^{-3}$ & 737.34 $\pm$ 8.22 & 26.482 $\pm$ 0.159 \\
		& & $1.5$ & $7.251 \times 10^{-4}$ & 64.3 $\pm$ 2.7 & 6.360 $\pm$ 0.196 \\
		& & $1.9$ & $7.716 \times 10^{-4}$ & 69.50 $\pm$ 2.71  & 6.71 $\pm$ 0.19   \\ \cmidrule{2-6}
		& \multirow{3}{*}{$3~TeV$}  & $1.1$ & $5.060 \times 10^{-4}$ & 44.15 $\pm$ 2.37 & 4.875 $\pm$ 0.202 \\
		& & $1.5$ & $4.716 \times 10^{-4}$ & 41.23 $\pm$ 2.28 & 4.6 $\pm$ 0.2 \\
		& & $1.9$ & $4.713 \times 10^{-4}$ & 41.13 $\pm$ 2.29 & 4.627 $\pm$ 0.201 \\ \midrule
		
		\multirow{6}{*}{$pp \rightarrow 4l$} & \multirow{3}{*}{$1~TeV$}  & $1.1$ & $6.310 \times 10^{-4}$ & 55.27 $\pm$ 2.62 & 7.382 $\pm$ 0.178 \\
		& & $1.5$ & $4.422 \times 10^{-5}$ &  3.82 $\pm$ 0.72 & 1.778 $\pm$ 0.202 \\
		& & $1.9$ & $5.903 \times 10^{-5}$ & 5.222 $\pm$ 0.776 & 2.128 $\pm$ 0.184 \\ \cmidrule{2-6}
		& \multirow{3}{*}{$3~TeV$}  & $1.1$ & $1.304 \times 10^{-5}$ & 1.051 $\pm$ 0.452 & 0.773 $\pm$ 0.243 \\
		& & $1.5$ & $1.026 \times 10^{-5}$ & 0.810 $\pm$ 0.413 & 0.639 $\pm$ 0.248 \\
		& & $1.9$ & $1.021 \times 10^{-5}$ & 0.806 $\pm$ 0.412 & 0.636 $\pm$ 0.248 \\ 
		\hline\hline
	\end{tabular}
\end{table}	

In this part, we discuss the signal with two isolated electrons and two isolated muons at the LHC. The other possibilities, that is, four electrons or four muons, show very similar features. Even though four lepton isolation is considered to be a difficult signal to pursue, we nonetheless explore it here as a case study with the results depicted in Figs.~\ref{var2e2m} and \ref{pt2e2m}. We could conclude that the signal shows some order of magnitude deviations from the background as long as $d$ is small like 1.1 or so. The deviation is there even for $\Lambda_{\cal U}=3$ TeV. However, as we allow $d$ to be larger, only for $\lambda=1$ TeV case shows profound differences from the background and as $\Lambda_{\cal U}$ gets larger the signal goes below the background where the signal identification would require new techniques.


To summarize the situation and to be able to compare roughly the signals with each other, it would be useful to calculate the significance of each signal, defined as $S/\sqrt{S+B}$ where $S(B)$ is the number of expected Signal (Background) events. Then any signal with significance larger than $5$ and having at least 5 signal events could be qualified as a potential venue for tracing new physics effects. Our results are summarized in Table \ref{tab:significance}. The numeric results are generated with the use of {\tt MadAnalysis} \cite{Dumont:2014tja,*Conte:2014zja,*Conte:2012fm}
	
	 As seen from the table that the largest significance is for the $pp\to 2\gamma 2g$ case with lots of signal events. It should be noted that $2\gamma 2 g$ signal may not be easy to detect due to gluon jet involvement. For the $pp\to 4\gamma$ case, the unparticle effects are sizable only  for $\Lambda_{\cal U}=1$ TeV with the scaling parameter $d$ near to its boundary values. As compared to the $4\gamma$ case, the situation in the $pp\to 2\gamma 2\ell$ case is similar but for even $\Lambda_{\cal U}=3$ TeV the significance is very close to 5. The background for the $pp\to 4\ell$ is large enough so that the unparticle effects may have a chance to be distinguishable for only $(d_{\cal U},\Lambda_{\cal U})=(1.1,1\,{\rm TeV})$. It should also be noted that the signal over background ratio can be enhanced by doing a further cut optimization, which can be deduced from the distributions shown.

\section{Conclusion\label{Sec:conc}}
	
If a scale invariant sector exists and finds ways to interact with the SM fields through heavy mediators, the scenario with some scalar, vector or tensor unparticle has been realized, and the possibility that the unparticle is indeed a scalar seems to be phenomenologically favored. 

In this study signals with final states; 4 photons, 2 photons + 2 gluons, 2 photons + 2 leptons, 2 electrons + 2 muons in the proton-proton collisions at the LHC at 14 TeV center of mass energy are considered within the framework of the scalar unparticle scenario after implementing the three-point self-interactions of the scalar unparticles in ${\tt MadGraph}$ while keeping all other possible contributions to the signals. 

We first discuss possible bounds on the parameters of the model from the available  $pp\to 2\gamma$ analysis with the data at 7 TeV \cite{Chatrchyan:2012xdj}. The current bounds for $\lambdau$ are around $1\,$TeV and hence unparticle signals are expected to be seen at LHC. The signals mentioned above are discussed after putting some basic cuts and compared with the SM predictions. The number of events with integrated luminosity $100$ fb$^{-1}$ as a function of various quantities like number of jets, $H_T$, $\Delta R_{ij}$, $m_{ij}$, and $p_T$ are depicted. We also summarize the results together with the significance of each signal in Table \ref{tab:significance}. It seems that indirect unparticle effect could be discriminated from the SM in almost all cases if the cut-off scale $\Lambda_{\cal U}$ is around 1 TeV and for especially small $d_{\cal U}$ values, close to its lower boundary value and, in some cases, to its upper boundary value as well. Taking into account the increase in  the precision achieved in the ongoing experiments, it is conceivable that experiments could probe to discriminate the unparticle effects at LHC in near future.

\acknowledgments
We are much obliged to late beloved Prof. Nam{\i}k Kemal Pak for him being a role model to us, father figure to some of us, his constant guidance and endless motivation as well as being an ideal colleague. We are deeply saddened by his sudden and untimely loss. We dedicate this study to his memory.     

I.T. thanks TUBA-GEBIP for its partial support. We thank Olivier Mattelaer for his help to implement the model in {\tt MadGraph}.

\clearpage
\appendix
\section{Feynman Diagrams Contributing to the Processes}\label{fdlist}
In this part, instead of presenting the complete list of the Feynman diagrams contributing to the considered processes, we prefer to give sample of diagrams corresponding different topologies in each case.
\begin{figure}[htb]
	\includegraphics[scale=0.7]{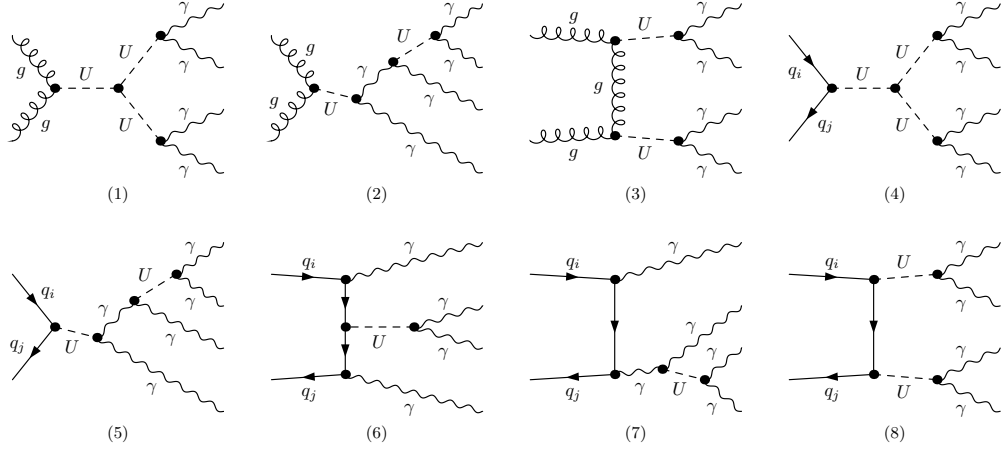}   
	\caption{Selected Feynman diagrams for the process $pp\to 4\gamma$. All possible permutations should be added to get the full list.  }
\end{figure}
\begin{figure}[htb]
	\includegraphics[width=14cm,height=9cm]{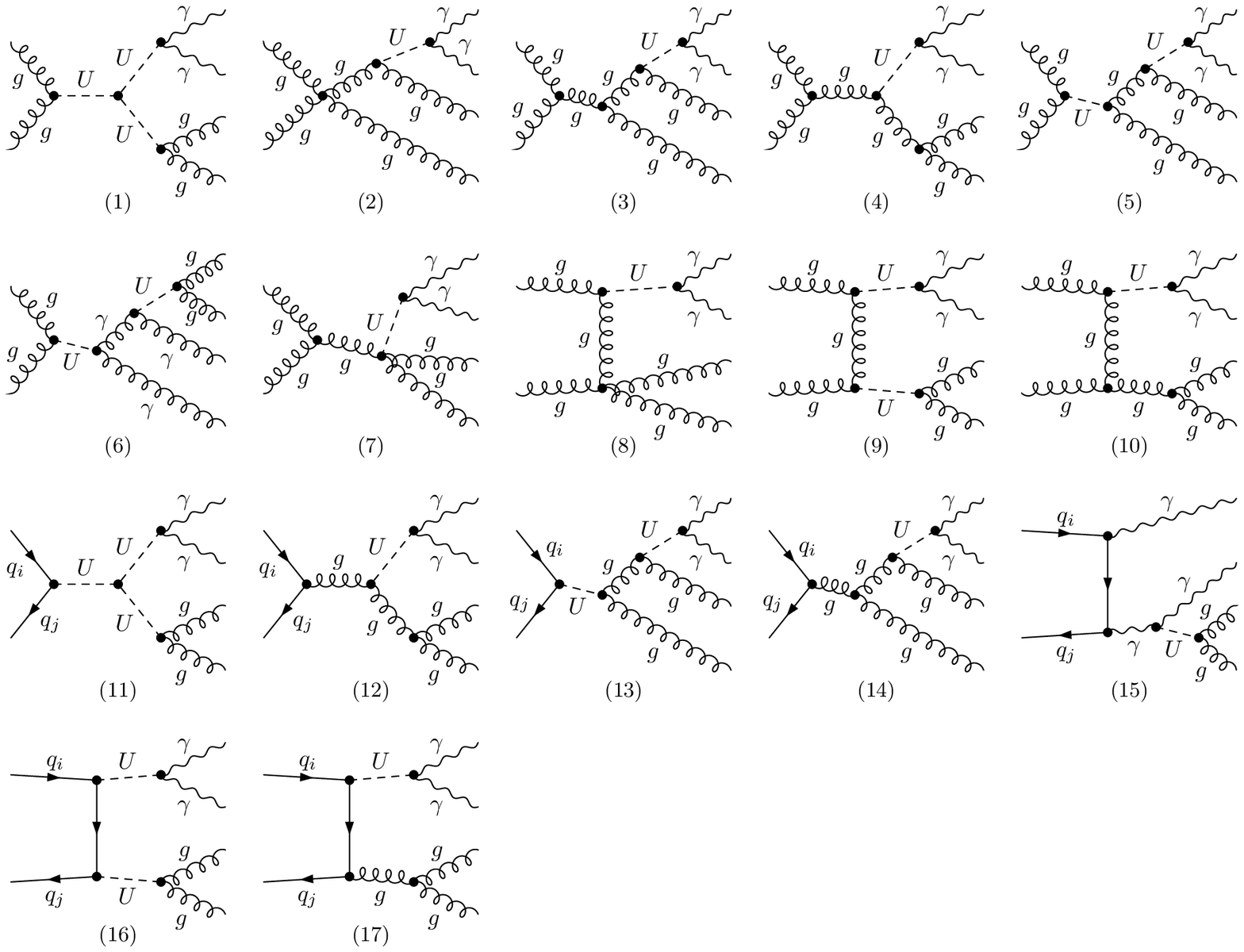}   
	\caption{Selected Feynman diagrams for the process $pp\to 2\gamma 2g$. All possible permutations should be added to get the full list.  }
	\vskip -0.9cm
\end{figure}
\begin{figure}[htb]
	\includegraphics[scale=0.8]{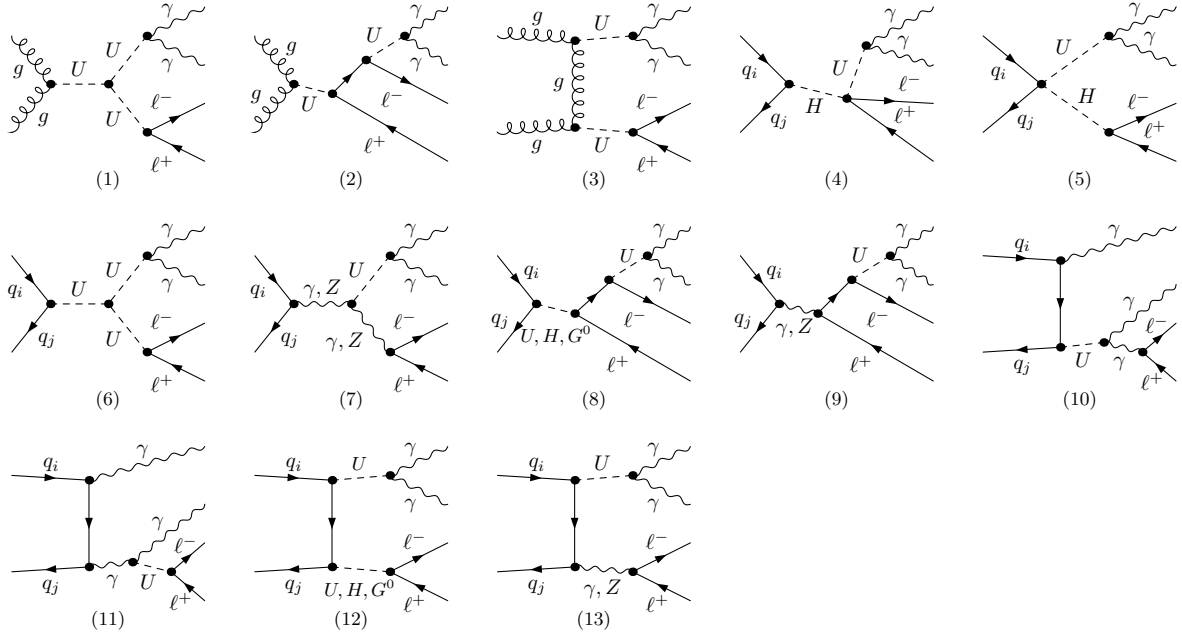}   
	\caption{Selected Feynman diagrams for the process $pp\to 2\gamma \ell^+\ell^-$. Here $\ell=e,\mu$. All possible permutations should be added to get the full list. }
	\label{fig:feyn_pp2p2l}
\end{figure}
\begin{figure}[htb]
	\includegraphics[scale=0.9]{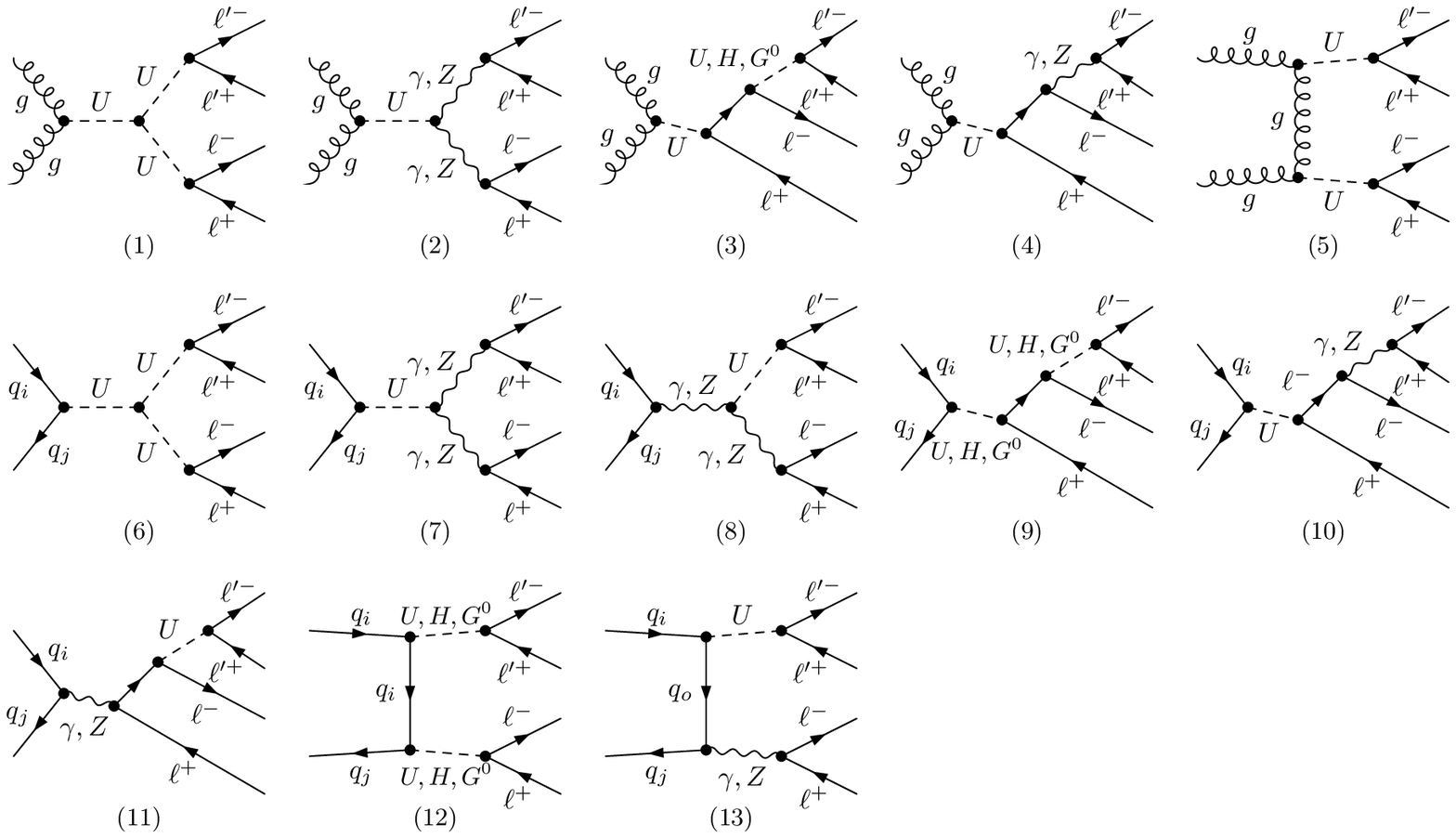}   
	\caption{Selected Feynman diagrams for the process $pp\to \ell^+\ell^-\ell^{\prime +} \ell^{\prime -}$. Here $\ell,\ell^\prime=e,\mu$. All possible permutations should be added to get the full list.}
	\label{fig:feyn_pp4l}
\end{figure}

\clearpage
\section{Polynomial Forms of the Three Point Correlation Function}\label{func}

Here we list, for various $d_{\cal U}$ values, the fitted form of the $F_y$ function which is written as:
\begin{eqnarray}
F_y \left(\frac{p_1^2}{s}, \frac{p_2^2}{s} \right) = 10^{f_y\left(\log_{10} \frac{p_1^2}{s},\log_{10} \frac{p_2^2}{s},\,d_{\cal U}\right)}
\end{eqnarray}
where the function $f_y(x, y, d_{\cal U})$ is assumed to be a sixth order polynomial with variables $x$ and $y$.

The explicit form of $f_y(x, y, d_{\cal U})$ for each $d_{\cal U}$ values are listed below:

\begin{eqnarray}
f_y(x, y, 1.1) &=& -1.5405271000000-0.7774200200000x-0.7866330300000y\nonumber \\
&&+0.0792432390000 x^2-0.0568065580000xy+0.0536183890000y^2 \nonumber\\
&&+0.0323167350000x^3-0.0257729210000x^2y-0.0178391900000xy^2 \nonumber\\
&&+0.0113794130000y^3 + 0.0088208374000x^4-0.0091290374000x^3y \nonumber\\
&&+0.0013427708000x^2y^2-0.0091546736000xy^3+0.0028157322000y^4 \nonumber\\
&&+0.0013727963000x^5-0.0014610970000x^4y -0.0003212246200x^3y^2 \nonumber\\
&&+0.0008116610800x^2y^3-0.0022305525000xy^4+0.0006806753400y^5 \nonumber\\
&&+0.0000870754330x^6-0.0000583811290x^5y -0.0002241075200x^4y^2 \nonumber\\
&&+0.0004523071200x^3y^3-0.0004087510700x^2y^4+0.0000568526710xy^5  \nonumber\\
&&+0.0000170296180y^6, \nonumber \\
f_y(x,y,1.2) &=& -1.5931413000000-0.7289910400000x-0.7365683400000y \nonumber\\
&&+0.0495076420000x^2-0.0332335900000xy+0.0276983440000y^2 \nonumber\\
&&+0.0270191700000x^3-0.0202016920000x^2y-0.0104586380000xy^2 \nonumber\\
&&+0.0084081501000y^3+0.0100315200000x^4-0.0106389120000x^3y \nonumber\\
&&+0.0041248972000x^2y^2-0.0080704982000xy^3+0.0040419030000y^4 \nonumber\\
&&+0.0019167613000x^5-0.0022792017000x^4y+0.0004435513600x^3y^2 \nonumber\\
&&+0.0005678533700x^2y^3-0.0017910916000xy^4+0.0009702022600y^5 \nonumber\\
&&+0.0001383103600x^6-0.0001455056900x^5y-0.0001537687100x^4y^2 \nonumber\\
&&+0.0004309380400x^3y^3-0.0004127246000x^2y^4+0.0000962308500xy^5 \nonumber\\
&&+0.0000340802980y^6, \nonumber\\
f_y(x,y,1.3) &=& -1.6410233000000-0.6831394500000x-0.6839238600000y \nonumber\\
&&+0.0098356018000x^2-0.0061022750000xy+0.0094178007000y^2 \nonumber\\
&&+0.0060856850000x^3-0.0046300120000x^2y-0.0048812233000xy^2 \nonumber\\
&&+0.0092921323000y^3+0.0028227822000x^4-0.0043816065000x^3y \nonumber\\
&&+0.0042084365000x^2y^2-0.0063283361000xy^3+0.0059037593000y^4 \nonumber\\
&&+0.0006131535500x^5-0.0010425094000x^4y+0.0005225122400x^3y^2 \nonumber\\
&&+0.0003664327100x^2y^3-0.0013713355000xy^4+0.0013363554000y^5 \nonumber\\
&&+0.0000470772780x^6-0.0000616165740x^5y-0.0000958230540x^4y^2 \nonumber\\
&&+0.0003070295000x^3y^3-0.0003082150200x^2y^4+0.0000764330580xy^5 \nonumber\\
&&+0.0000649171240y^6, \nonumber\\
f_y(x,y,1.4) &=& -1.6842770000000-0.6268423200000x-0.6338316600000y \nonumber\\
&&+0.0055970389000x^2+0.0139964740000xy-0.0145320950000y^2 \nonumber\\
&&+0.0192604490000x^3-0.0026778417000x^2y+0.0034019656000xy^2 \nonumber\\
&&-0.0004070407400y^3+0.0102869040000x^4-0.0068507861000x^3y \nonumber\\
&&+0.0059080722000x^2y^2-0.0031578032000xy^3+0.0017214913000y^4 \nonumber\\
&&+0.0021861777000x^5-0.0020625485000x^4y+0.0013463745000x^3y^2 \nonumber\\
&&-0.0001467668100x^2y^3-0.0003119999900xy^4+0.0002904937700y^5 \nonumber\\
&&+0.0001643119500x^6-0.0001700669500x^5y+0.0000040501655x^4y^2 \nonumber\\
&&+0.0002493650600x^3y^3-0.0003028055500x^2y^4+0.0001574897700xy^5 \nonumber\\
&&-0.0000238051790y^6, \nonumber\\
f_y(x,y,1.5) &=& -1.7239823000000-0.5841207000000x-0.5835451200000y \nonumber\\
&&-0.0329579040000x^2+0.0277305850000xy-0.0249275450000y^2 \nonumber\\
&&-0.0061406726000x^3+0.0071845036000x^2y+0.0009698820400xy^2 \nonumber\\
&&+0.0067371962000y^3-0.0005908533400x^4+0.0009089441600x^3y \nonumber\\
&&+0.0018573577000x^2y^2-0.0039214458000xy^3+0.0065187567000y^4 \nonumber\\
&&-0.0000746652690x^5+0.0003873853700x^4y-0.0003799170600x^3y^2 \nonumber\\
&&+0.0003096119100x^2y^3-0.0008651676500xy^4+0.0014011978000y^5\nonumber\\
&&-0.0000091955711x^6+0.0000704813420x^5y-0.0001640648600x^4y^2 \nonumber\\
&&+0.0002406092800x^3y^3-0.0002385580800x^2y^4+0.0000744175370xy^5 \nonumber \\
&&+0.0000706166990y^6, \nonumber \\
f_y(x,y,d_1.6) &=& -1.7574221000000-0.5296548400000x-0.5320717200000y \nonumber \\ &&
-0.0381500120000x^2+0.0478045870000xy-0.0410908730000y^2  \nonumber\\  &&
+0.0011269690000x^3+0.0108900100000x^2y+0.0084243263000xy^2 \nonumber\\ &&
+0.0004794936600y^3+0.0036834885000x^4-0.0011481639000x^3y \nonumber\\ &&
+0.0049334989000x^2y^2-0.0018404656000xy^3+0.0035994200000y^4 \nonumber\\ &&
+0.0008605367100x^5-0.0006515736200x^4y+0.0008031427900x^3y^2) \nonumber\\ &&
-0.0000690085460x^2y^3-0.0001479700300xy^4+0.0006790255900y^5 \nonumber\\ &&
+0.0000633053100x^6-0.0000456439360x^5y-0.0000330677610x^4y^2 \nonumber\\ &&
+0.0001861850800x^3y^3-0.0002131552700x^2y^4+0.0001142648200xy^5 \nonumber\\ &&
+0.0000146061000y^6, \nonumber \\
f_y(x,y,d_1.7) &=& -1.7859310000000-0.4819532800000x-0.4808190400000y \nonumber\\ &&
-0.0568205160000x^2+0.0601820280000xy-0.0510483810000y^2 \nonumber\\ &&
-0.0085626578000x^3+0.0157293060000x^2y+0.0118361580000xy^2 \nonumber\\ &&
-0.0015692444000y^3-0.0003973136300x^4+0.0020748123000x^3y \nonumber\\ &&
+0.0033761192000x^2y^2+0.0002446355700xy^3+0.0026363003000y^4 \nonumber\\ &&
+0.0000200407320x^5+0.0002556639700x^4y+0.0002493241300x^3y^2 \nonumber\\ &&
+0.0001335396100x^2y^3+0.0000923550460xy^4+0.0004898155200y^5 \nonumber\\ &&
-0.0000000764952x^6+0.0000324266770x^5y-0.0000549079190x^4y^2 \nonumber\\ &&
+0.0001259923600x^3y^3-0.0001257337300x^2y^4+0.0000846048940xy^5 \nonumber\\ &&
+0.0000097441803y^6, \nonumber \\
f_y(x,y,1.8) &=& -1.8082217000000-0.4311959700000x-0.4321981400000y \nonumber\\ &&
-0.0614003580000x^2+0.0681025750000xy-0.0630279150000y^2 \nonumber\\ &&
-0.0052333005000x^3+0.0147989660000x^2y+0.0142681270000xy^2 \nonumber\\ &&
-0.0062819835000y^3+0.0015725778000x^4+0.0005401446100x^3y \nonumber\\ &&
+0.0041388541000x^2y^2+0.0004103011300xy^3+0.0010780395000y^4 \nonumber\\ &&
+0.0004473018100x^5-0.0001973343900x^4y+0.0004299858200x^3y^2 \nonumber\\ &&
+0.0002349362300x^2y^3-0.0000267569530xy^4+0.0002615601200y^5 \nonumber\\ &&
+0.0000327805430x^6-0.0000127250450x^5y-0.0000178441340x^4y^2 \nonumber\\ &&
+0.0000848834490x^3y^3-0.0000701021930x^2y^4+0.0000425878570xy^5 \nonumber\\ &&
+0.0000031287267y^6, \nonumber \\ 
f_y(x,y,1.9) &=& -1.8230678000000-0.3813089500000x-0.3805333500000y \nonumber\\ &&
-0.0644029630000x^2+0.0692731770000xy-0.0617737870000y^2 \nonumber\\ &&
-0.0037091815000x^3+0.0127173990000x^2y+0.0135749080000xy^2 \nonumber\\ &&
-0.0010125714000y^3+0.0023442925000x^4-0.0000228801960x^3y \nonumber\\ &&
+0.0028946364000x^2y^2+0.0005196745700xy^3+0.0034504318000y^4 \nonumber\\ &&
+0.0005942434900x^5-0.0002497974400x^4y+0.0001526318300x^3y^2 \nonumber\\ &&
+0.0002040476400x^2y^3-0.0000738331010xy^4+0.0007605124000y^5 \nonumber\\ &&
+0.0000429606120x^6-0.0000158586280x^5y-0.0000256402560x^4y^2 \nonumber\\ &&
+0.0000515125370x^3y^3-0.0000392382300x^2y^4+0.0000179126770xy^5 \nonumber\\ &&
+0.0000458076720y^6. 
\end{eqnarray}

\bibliography{unparticle.bib}{}

\end{document}